\documentclass[fleqn,usenatbib]{mnras}

\usepackage{newtxtext,newtxmath}

\usepackage[T1]{fontenc}

\DeclareRobustCommand{\VAN}[3]{#2}
\let\VANthebibliography\thebibliography
\def\thebibliography{\DeclareRobustCommand{\VAN}[3]{##3}\VANthebibliography}

\usepackage{graphicx}	

\title[GALAH age-metallicity distributions]{Characterizing epochs of star formation across the Milky Way disc using age-metallicity distributions of GALAH stars}

\author[C. L. Sahlholdt et al.]{
Christian L. Sahlholdt$^{1}$\thanks{E-mail: christian.sahlholdt@astro.lu.se; sofia.feltzing@astro.lu.se},
Sofia Feltzing$^{1}$\footnotemark[1]
and Diane K. Feuillet$^{1}$
\\
$^{1}$Lund Observatory, Department of Astronomy and Theoretical Physics, Box 43,
      SE-221 00 Lund, Sweden\\
}

\date{Accepted XXX. Received YYY; in original form ZZZ}

\pubyear{2021}

\begin{document}
\label{firstpage}
\pagerange{\pageref{firstpage}--\pageref{lastpage}}
\maketitle

\begin{abstract}
We provide a detailed map of the ages and metallicities of turnoff stars in the Milky Way disc based on data from GALAH DR3 and \textit{Gaia} EDR3.
From this map, we identify previously undetected features in the age-metallicity distribution of disc stars and interpret these results as indicating a three-phase formation history of the Milky Way.
In the first phase, inner disc stars form along a single age-metallicity sequence and are today kinematically hot.
The end of this phase is marked by a local minimum in the inner disc age distribution 10~Gyr ago.
At this time, we find the stellar populations to transition from high to low alpha-element abundances and from high to low vertical velocity dispersion.
In the second phase, stars form across the disc with outwardly decreasing metallicity.
In this phase, inner disc stars form at super-solar metallicites in a continuation of the early age-metallicity relation while outer disc stars begin forming at metallicities at least 0.5~dex lower.
Finally, the third phase is associated with a recent burst of star formation across the local disc marked by a local minimum in the age-metallicity distribution 4 to 6~Gyr ago.
Future quantitative comparisons between the observed age-metallicity distribution and those of simulated galaxies could help constrain the processes driving each of the star formation phases.
\end{abstract}

\begin{keywords}
Galaxy: stellar content -- Galaxy: disc -- Galaxy: formation -- Galaxy: evolution
\end{keywords}



\section{Introduction}

The formation and evolution of the Milky Way is one of the main research topics in modern astrophysics.
In the Milky Way, individual stars can be characterized in great detail.
They carry information about the formation history of the Galaxy in their ages, elemental abundances, and kinematics.
In recent years, these important parameters have been obtained for large samples of stars in the solar neighbourhood and beyond.
This is exemplified by the \textit{Gaia} mission \citep{2016A&A...595A...1G} which has provided extremely precise astrometric data for more than a billion stars.
At the same time, elemental abundances for several millions of stars are provided by large spectroscopic surveys such as the Galactic Archaeology with HERMES \citep[GALAH;][]{2015MNRAS.449.2604D} survey.
The combination of \textit{Gaia} and spectroscopic data also allow for stellar ages to be estimated e.g. by fitting stellar evolution models to the data.
These large samples of stars characterized by the data from major surveys can be used to constrain models of Milky Way formation and aid our understanding of galaxy formation in general.

One of the observational constraints on models of Galaxy formation is the relationship between stellar age and metallicity.
The age-metallicity relation has been an object of study since some of the earliest investigations into the chemical evolution of the Galactic disc.
Perhaps the most influential of such early investigations was carried out by \citet{1993A&A...275..101E} who found no strong age-metallicity relation in the solar neighborhood.
Instead, the relation was found to be nearly flat with a large scatter in metallicity at all ages, beyond the observational uncertainties, except for a lack of old high-metallicity stars.
These trends have since been found in several other studies of nearby stars \citep{2001A&A...377..911F, 2011A&A...530A.138C, 2013A&A...560A.109H, 2014A&A...565A..89B}.
\citet{1993A&A...275..101E} interpreted the metallicity scatter at a given age as an indication that these stars were born in different regions of the Galaxy with distinct chemical enrichment histories.
This is still the prevailing interpretation and spatial variations in the age-metallicity relation have since been observed directly in stellar samples covering a larger volume of the Galactic disc \citep{2019MNRAS.489.1742F, 2019ApJ...871..181H, 2021ApJ...920...23J}.

Additional scatter in the age-metallicity relation is caused by uncertainties on the age estimates.
Only in special cases, for example using asteroseismology \citep{2014A&A...569A..21L, 2018MNRAS.475.5487S} or eclipsing binaries \citep{2009AJ....137.5086M, 2015A&A...579A..59V}, can individual stellar ages be determined with a relative precision of 10 per cent or better.
This level of precision is not attainable for the bulk of stars observed as part of spectroscopic surveys.
For these stars, age estimates based on stellar model fitting are usually limited to a relative precision of 20 per cent \citep{2010ARA&A..48..581S}.
Recently developed machine learning methods used to infer ages of giants based on spectral features yield ages with relative precisions of 30 to 40 per cent \citep{2016MNRAS.456.3655M, 2019MNRAS.484.5315W}.
These age uncertainties act to smear out the age-metallicity distribution, especially among the oldest stars.
Therefore, age uncertainties is one of the main obstacles in the search for features in the age-metallicity distribution predicted in models of Milky Way formation \citep[see e.g.][their Fig. 12]{2021MNRAS.503.5846R}.
Other obstacles to detailed comparison between models and observations include our ability to analyse stellar spectra with correct physical detail at all steps.
For example, the neglect of non-local thermodynamic equilibrium can in some cases result in very different elemental abundance trends \citep{2020A&A...642A..62A}.

Statistical methods aimed at inferring the age distribution of a stellar sample without assigning each star an individual age may provide a more precise recovery of the underlying age distribution.
This can be achieved by combining the full age probability density functions of individual stars calculated based on stellar model fitting.
Such methods have previously been shown to give a better recovery of the underlying age distribution of synthetic stellar samples than the distribution of individual age estimates \citep{1999MNRAS.304..705H, 2005ESASP.576..171J}.
As highlighted by \citet{2019A&A...629A.127M}, this also accounts for potential non-Gaussian or multi-modal age probability density functions which would otherwise be reduced to a potentially misleading point estimate.
Recently, \citet{2021MNRAS.502..845S} presented a method which extends this idea to age-metallicity distributions.
With this method a more precise age-metallicity distribution can be obtained when large stellar samples are available.

In this work, we present the first large-scale application of the method presented by \citet{2021MNRAS.502..845S} to infer the age-metallicity distribution of dwarf and subgiant stars in the GALAH survey.
Thanks to a sample size of roughly 180\,000 stars, we are also able to break the distribution down according to stellar kinematics and positions in the Galactic disc.
These distributions reveal features which are not usually found in age-metallicity scatter plots based on individual age estimates.
For example, we find local minima in the distribution at ages around 5 and 10~Gyr as well as a clear age-metallicity relation for the oldest stars in the inner Galactic disc.

In Section~\ref{sec:data_and_methods} the sample selection is described and the method used to estimate age-metallicity distributions is summarised.
The results are presented in Section~\ref{sec:results}, first for the sample as a whole and then for kinematically and spatially selected subsamples.
A discussion of the results and their relation to models of Milky Way formation is given in Section~\ref{sec:discussion}.
Finally, our conclusions are summarised in Section~\ref{sec:conclusions}.

\section{Data and methods} \label{sec:data_and_methods}

Throughout this work we use spectroscopic stellar parameters from the Third Data Release of GALAH \citep[DR3,][]{2021MNRAS.506..150B} combined with parallaxes from \textit{Gaia} Early Data Release 3 \citep[EDR3,][]{2021A&A...649A...1G}.
We also make use of Galactocentric positions, velocities, and orbital parameters from the GALAH DR3 value-added catalogue\footnote{For this work we have used the value-added catalogue (VAC) version 2. Both the main catalogue and the VAC were downloaded from \url{https://www.galah-survey.org/dr3/overview/} (Accessed December 2021).} which is based on astrometry from \textit{Gaia} EDR3 and radial velocities primarily determined from the GALAH spectra \citep{2020arXiv201212201Z}.
The orbital parameters in this catalogue, such as the maximum distance from the Galactic disc plane $z_{\mathrm{max}}$, have been calculated using the Python package \texttt{galpy} \citep{2015ApJS..216...29B}.
Details about the assumed Milky Way potential and solar kinematic parameters can be found in the GALAH release paper \citep{2021MNRAS.506..150B}.

\subsection{Sample selection}

\begin{figure}
\centering
\includegraphics[width=\columnwidth]{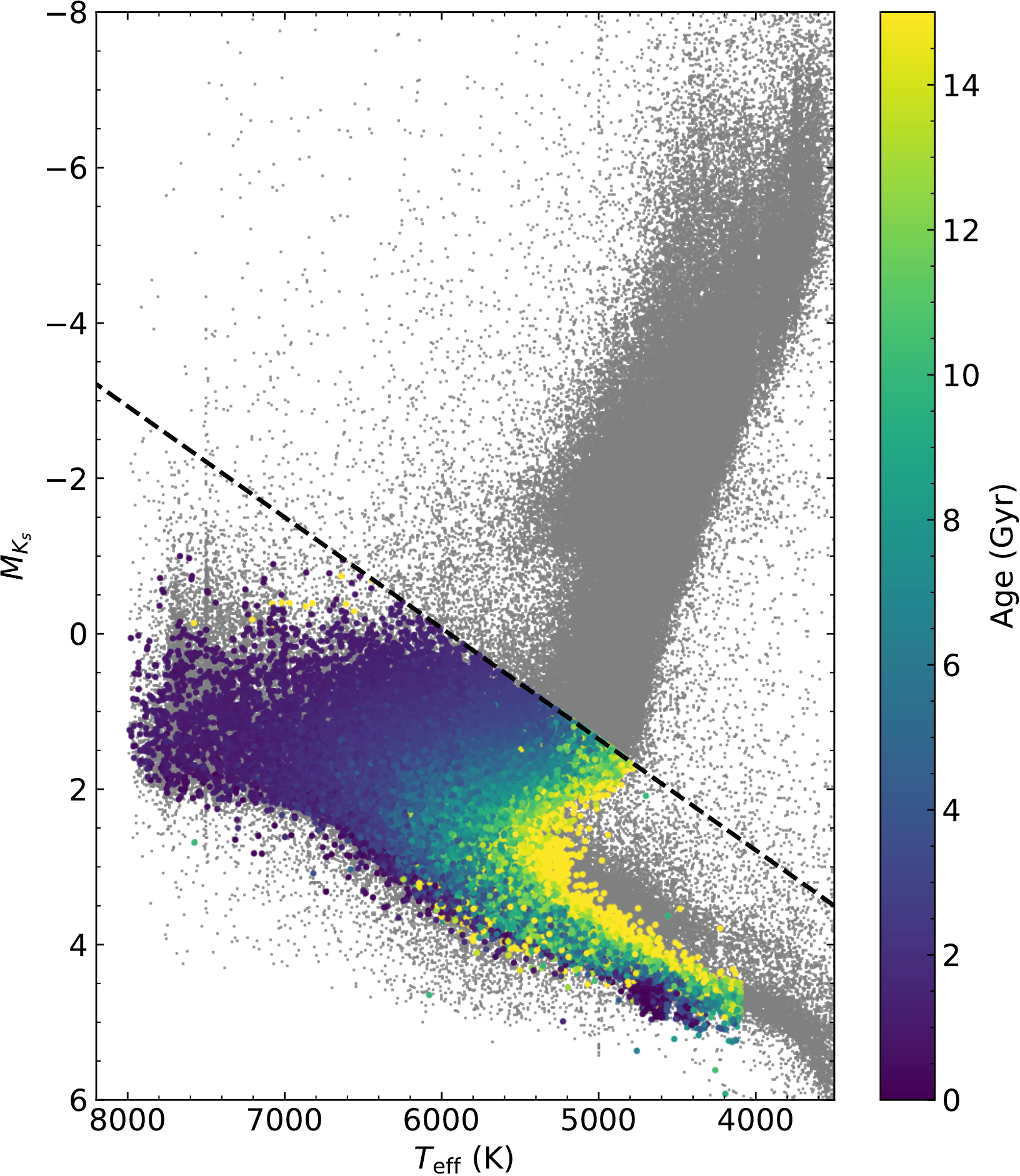}
\caption{HR diagram of all stars in GALAH DR3 (grey) and the selected sample coloured by individual age estimates from this work. The black dashed line indicates the cut made to exclude giant stars (see Eq.~\eqref{eq:magnitude_cut}).}
\label{fig:HRD_ages}
\end{figure}

Only stars observed as part of the main GALAH survey (\texttt{survey\_name~=~galah\_main} in the catalogue) are considered in this study.
Following the recommendations in GALAH DR3 we use the stellar parameter flags in the catalogue to make quality cuts.
The sample is limited to the stars with \texttt{flag\_sp~=~0}, \texttt{flag\_fe\_h~=~0}, and \texttt{flag\_alpha\_fe~=~0}, which indicate that no problems were identified with the stellar parameter determination including iron and alpha-element abundances.
Limiting the sample to \texttt{flag\_sp~=~0} also excludes stars identified as possible binaries and cuts out stars with $T_{\mathrm{eff}} \lesssim 4000$~K. 
Additionally, a single cut is made based on the \textit{Gaia} EDR3 parameters by selecting only stars with \texttt{ruwe~<~1.4}.\footnote{RUWE is the renormalised unit weight error and is a measure of the astrometric goodness-of-fit \citep{2021A&A...649A...2L}.}
No cut is made based on the uncertainty of the parallax, $\varpi$, since 99.95 per cent of the stars in the selected sample have a relative parallax uncertainty below 10 per cent.
Finally, we exclude giant stars with the following cut on the absolute magnitude:
\begin{equation} \label{eq:magnitude_cut}
    M_{K_{s}} = m_{K_{s}}-A_{K_{s}}-5\log10[(100~\mathrm{mas})/\varpi] > 8.5-T_{\mathrm{eff}}/(700~\mathrm{K}) \; .
\end{equation}
The 2MASS $m_{K_{s}}$ magnitudes \citep{2006AJ....131.1163S} and the extinction values $A_{K_{s}}$ are taken from the GALAH catalogue.
The cut in Eq.~\eqref{eq:magnitude_cut} was made by eye in the Hertzsprung--Russell (HR) diagram (\autoref{fig:HRD_ages}) to remove the giant branch while retaining the hottest turnoff stars.

The final selection, cut down from the roughly 600\,000 stars in GALAH DR3, contains about 180\,000 dwarfs and subgiants in total.

\subsection{Stellar model fitting}

The estimation of age-metallicity distributions begins with fitting a grid of stellar evolution models to each individual star to determine their two-dimensional (2D) $\mathcal{G}$~functions, $\mathcal{G}(\tau, \zeta|\mathbf{x})$, following \citet{2019A&A...622A..27H}.
The function $\mathcal{G}(\tau, \zeta|\mathbf{x})$ is the joint marginal likelihood of the age, $\tau$, and initial metallicity, $\zeta$, given a set of observables, $\mathbf{x}$.

\begin{table}
    \caption{Ranges of ages and initial metallicites for the two grids of stellar models.
    The indicated step-size gives the grid resolution in each parameter.
    The low-resolution grid has a wider range of parameter values but only half the resolution.}
    \begin{tabular}{lrrrrrrr}
    \multicolumn{1}{c}{} & \multicolumn{3}{c}{Main grid} &  & \multicolumn{3}{l}{Low-resolution grid} \\ \cline{2-4} \cline{6-8} 
                         & Min      & Max     & Step     &  & Min          & Max        & Step        \\ \hline
    Age (Gyr)            & $0.1$        & $15.0$      & $0.2$      &  & $0.1$            & $18.0$         & $0.4$         \\
    $[\mathrm{M}/\mathrm{H}]_{\mathrm{initial}}$          & $-0.70$     & $0.50$     & $0.05$     &  & $-1.50$         & $0.50$        & $0.10$    \\ \hline    
    \end{tabular}
    \label{tab:model_grids}
\end{table}

We use a grid of PARSEC stellar evolution models \citep{2012MNRAS.427..127B} with each model described by its age, $\tau$, initial metallicity, $\zeta = [\mathrm{M}/\mathrm{H}]_{\mathrm{initial}}$, initial mass, $m$, and distance modulus, $\mu$.
The distance modulus is added as an external parameter which determines the parallax and apparent magnitude of the model.
The grid contains models from $0.1$ to $15$~Gyr in age, with a resolution of $0.2$~Gyr, and initial metallicities from $-0.7$ to $0.5$~dex, with a resolution of $0.05$~dex.
The limited range of metallicities is chosen because the computation time of the age-metallicity distribution (see Section~\ref{sec:method_SAMD}) increases rapidly with the number of grid points.
We have chosen the lower metallicity limit of $-0.7$~dex to make sure that the majority of the sample (about 99 per cent) fall within the limits of the grid.
For the stars with lower metallicities we have defined a secondary model grid with lower resolution; the parameters for both grids are given in \autoref{tab:model_grids}.

For each model in the grid, the likelihood is calculated as
\begin{equation} \label{eq:isofit_likelihood}
    L(\tau, \zeta, m, \mu | \mathbf{x}) = \exp\left[ -\dfrac{1}{2} \sum_{i} \left( \dfrac{x_i-X_i(\tau, \zeta, m, \mu)}{\sigma_i} \right)^2 \right] \, ,
\end{equation}
where $X_i$ is the model parameter corresponding to the observed parameter $x_i$ with uncertainty $\sigma_i$.
The observables are $\mathbf{x}~=~(T_{\mathrm{eff}}, [\mathrm{M}/\mathrm{H}], m_{\mathrm{K_s}}, \varpi)$, i.e., the effective temperature, metallicity, apparent magnitude in the 2MASS $\mathrm{K_{s}}$~band (corrected for extinction), and the parallax.
Here we define the metallicity following the prescription by \citet{1993ApJ...414..580S}:
\begin{equation}
    [\mathrm{M}/\mathrm{H}] = [\mathrm{Fe}/\mathrm{H}]+\log(0.638\times 10^{[\alpha/\mathrm{Fe}]}+0.362) \, .
\end{equation}
This is to include the effect of alpha-enhancement which is not considered in the models where the metallicity is given by $\zeta = \log(Z/X)-\log(Z/X)_{\odot}$ with $(Z/X)_{\odot} = 0.0207$.
[Fe/H] and [$\alpha$/Fe] are taken from the GALAH catalog.
In GALAH DR3, [$\alpha$/Fe] is reported as the error-weighted combination of the measured abundances for select lines of Mg, Si, Ca, and Ti, see \citet{2021MNRAS.506..150B} for details.

It is important to consider the difference in the models between the initial metallicity, $\zeta$, and the surface metallicity at later times, $\zeta'(\tau, \zeta, m)$.
For turnoff stars, the surface metallicity is generally lower than the initial metallicity due to atomic diffusion \citep[e.g.,][]{2004ApJ...606..452M}.
If this is not taken into account the stellar age may be overestimated by up to 20 per cent \citep{2017ApJ...840...99D}.
Therefore, the observed metallicity, [M/H], is compared with the surface metallicity, $\zeta'$, in Eq.~\eqref{eq:isofit_likelihood}.
However, the likelihood function is still defined on a grid of initial metallicities which carries over into the age-metallicity distributions described below.
In the figures in the following sections we label the initial metallicity $[\mathrm{M}/\mathrm{H}]_{\mathrm{initial}}$ instead of $\zeta$ to improve readability.

Having calculated the likelihood distribution over the entire grid of models, the 2D $\mathcal{G}$~function is found by marginalizing over the mass and distance modulus with priors $\xi(m)$ and $\psi(\mu)$
\begin{equation}
    \mathcal{G}(\tau, \zeta | \mathbf{x}) = \int_{\mu}\int_{m}\xi(m)\psi(\mu)L(m, \tau, \zeta, \mu | \mathbf{x})\mathrm{d}m \mathrm{d}\mu \, .
\end{equation}
In this work $\xi(m)$ is a Salpeter initial mass function\footnote{The choice of initial mass function makes little difference in the limited range of masses of the present sample, i.e. $\sim 0.7$ to $1.5~M_{\odot}$.} and $\psi(\mu)$ is flat.
The full details of this operation can be found in \citet[][Appendix A]{2019A&A...622A..27H}.

By marginalizing the 2D $\mathcal{G}$~function over metallicity (with a flat prior), we get the one-dimensional (1D) $\mathcal{G}$~function, $\mathcal{G(\tau)}$.
The age of the star can be estimated from this function in different ways; here we choose to use the median (50th percentile) of the distribution.
In \autoref{fig:HRD_ages} the sample is coloured by ages estimated in this way for each star.

\subsection{Estimating the age-metallicity distribution} \label{sec:method_SAMD}

In addition to using the $\mathcal{G}$~functions to derive individual stellar ages, before marginalizing over metallicity, the 2D $\mathcal{G}$~functions can be combined to estimate the age-metallicity distribution of the sample following the method by \citet{2021MNRAS.502..845S}.
To summarise, the method is based on minimising the expression
\begin{align} \label{eq:SAMD_eq}
    \begin{split}
        -\ln L' = &- \sum_{i} \ln\left( \int \mathcal{G}_{i}(\tau, \zeta)\phi(\tau, \zeta)\mathrm{d}\tau\mathrm{d}\zeta \right) \\ {}&+ \alpha \int\left(s_{\tau}^2\dfrac{\partial^2\phi}{\partial\tau^2} + s_{\zeta}^2\dfrac{\partial^2\phi}{\partial\zeta^2} \right)^2 \mathrm{d}\tau\mathrm{d}\zeta \, ,
    \end{split}
\end{align}
in order to find the age-metallicity distribution, $\phi(\tau, \zeta)$, that best explains the $\mathcal{G}$~functions of the sample.
The solution is subject to regularization given by the second term which penalises distributions with large second derivatives.
This term determines the smoothness of the distribution and is regulated in strength by the parameter $\alpha$.
The parameters $s_{\tau}$ and $s_{\zeta}$ act to nondimensionalize the regularization term and also set the relative strength of regularization along the two dimensions.
In this work these parameters are set equal to the grid resolution, so $s_{\tau} = 0.2$~Gyr and $s_{\zeta} = 0.05$~dex for calculations using the main grid.
We refer to the age-metallicity distribution that successfully minimises Eq.~\eqref{eq:SAMD_eq}, $\hat\phi(\tau, \zeta)$, as the sample age-metallicity distribution (SAMD).
Additionally, we can marginalize this distribution over metallicity to get the sample age distribution (SAD), $\hat\phi(\tau)$.
The SAMD and SAD are always normalized so they integrate to unity.
The implementation of the method used in this work is publicly available in the form of a Python script \citep{samd_2020}.

One of the advantages of using the SAMD is that no star needs to be removed from the sample based on its age uncertainty.
This uncertainty is contained in the $\mathcal{G}$~function and taken into account when the SAMD is calculated.
Tests of the SAMD on synthetic samples of turnoff stars have shown that it gives a better match to the true age distribution than the histogram of individually estimated ages for smooth, step-wise, and bursty star formation histories \citep{2021MNRAS.502..845S}.
This is also true when using only the most precise individual ages, e.g. by removing stars with a relative age uncertainty greater than 20 per cent.
Another advantage is that the SAMD includes the age-metallicity correlations inherent in model fitting.

A disadvantage of using the SAMD is that the choice of regularization parameter, $\alpha$, is ambiguous.
\citet{2019A&A...629A.127M} developed a method, similar to the one used in this work, for estimating the age distribution of a stellar sample.
This method also relies on a regularization parameter, and they used a cross-validation technique to estimate the optimal value of it.
Essentially, they broke their sample into smaller parts and required that each part yield similar solutions.
The solutions may differ significantly if the regularization parameter is too low because then small-scale variations specific to a certain subsample may dominate the solution.
Inspired by this approach, we calculate the SAMD, for a given value of $\alpha$, for several samples of 50\,000 stars drawn randomly from the full sample of 180\,000 stars.
Then we calculate the mean and standard deviation, at each age-metallicity grid point, of the different random realizations of the SAMD (\autoref{fig:SAMD_mean_std}).
To determine whether the solutions obtained from the different random subsamples agree, we use two quantities.
The first is the total noise obtained by taking the sum of the standard deviation over all grid points.
The second is the mean divided by the standard deviation, or the signal-to-noise (S/N), at each grid point.
We choose the smallest value of $\alpha$ for which the total noise is 20 to 40 per cent of the noise at $\alpha = 0$, and for which the S/N is smooth and greater than 10 over most of the age-metallicity space.
Although somewhat arbitrary, this choice is informed by testing the method on a synthetic sample of turnoff stars (\autoref{fig:SAMD_mean_std_synth}).

In addition to calculating the SAMD for the full sample, we are interested in studying the SAMD for subsamples selected according to the stellar kinematics or Galactic positions.
These subsamples (defined in sections~\ref{sec:vT_slice} and \ref{sec:Rz_slice}) have different sizes, down to about 1000 stars, so the value of $\alpha$ used for the full sample cannot be assumed to be optimal for them.
Therefore, we calculate the optimal value of $\alpha$ for one of these subsamples (\autoref{fig:SAMD_mean_std_popb}) and apply the same value when calculating the SAMD for every other subsample.
Further details on the choice of $\alpha$ are given in Appendix~\ref{app:choose_alpha}.

\section{Results} \label{sec:results}

\begin{figure}
\centering
\includegraphics[width=\columnwidth]{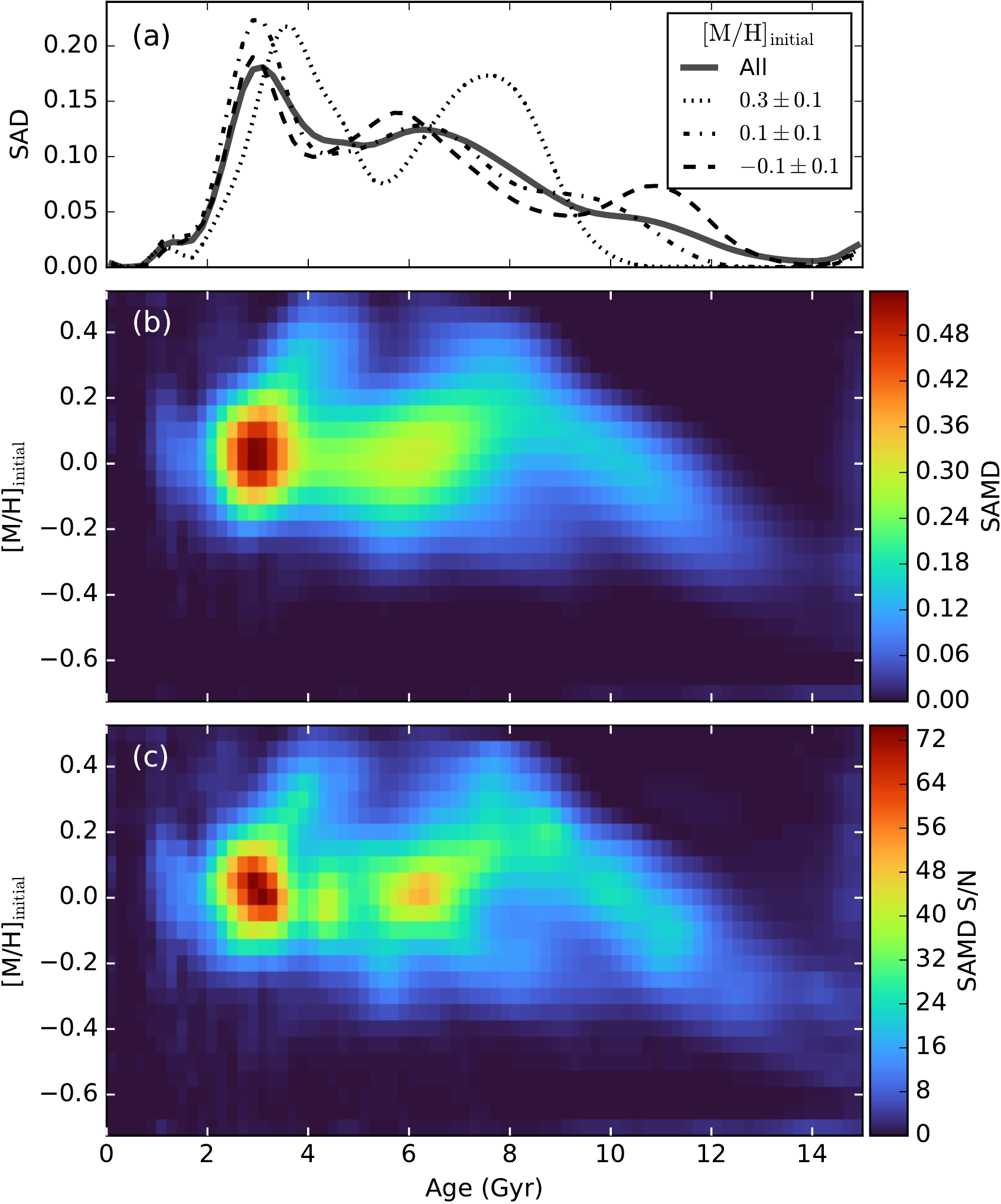}
\caption{(a) SAD for the full sample and for three intervals of initial metallicity as indicated in the legend.
These are calculated by selecting the part of the SAMD (panel b) in each metallicity interval and marginalizing over the metallicity.
(b) SAMD for the full sample of GALAH dwarf and subgiant stars.
This is the mean distribution resulting from running the method on 40 random draws of 50\,000 stars each.
(c) S/N of the SAMD calculated as the mean divided by the standard deviation of the 40 random draws of 50\,000 stars.}
\label{fig:SAMD_all}
\end{figure}

The SAMD and SAD of the full sample are shown in \autoref{fig:SAMD_all} along with the S/N of the SAMD calculated as described at the end of section~\ref{sec:method_SAMD}.
The SAMD shows some interesting structures; for example, there is a clear age-metallicity relation going from the oldest most metal-poor end of the distribution up to some of the highest metallicities in the sample at an age of 8~Gyr.
At lower ages most of the distribution is centered on $[\mathrm{M}/\mathrm{H}]_{\mathrm{initial}}= 0$~dex with a peak at ages between 2 and 4~Gyr.
From this peak there is also a ridge extending to higher metallicities at a slightly older age.

Previous studies of the age-metallicity relation in the Milky Way disc have found the distribution to be flat up to about 8~Gyr, decrease in metallicity above 8~Gyr, and show a large scatter in metallicity at all ages \citep[e.g.,][]{1993A&A...275..101E, 2001A&A...377..911F, 2011A&A...530A.138C, 2014A&A...565A..89B}.
These features are also present in \autoref{fig:SAMD_all}, but the SAMD reveals more structure in the age-metallicity distribution than seen in previous works.

The SAD of the full sample (\autoref{fig:SAMD_all}a) shows a local minimum at an age of about 5~Gyr and a shallow inflection at about 10~Gyr.
The location and strength of the minimum depends on the metallicity.
When considering the SAD for three different metallicity ranges, we find the following: At the highest metallicities ($[\mathrm{M}/\mathrm{H}]_{\mathrm{initial}} > 0.2$~dex; dotted line in the figure) the minimum at 5~Gyr is pronounced and the SAD shows two clear peaks.
There is an old peak at 8~Gyr, which is related to the young end of the old age-metallicity relation, and a young peak at 4~Gyr, which appears to be connected with the young peak at lower metallicities ($[\mathrm{M}/\mathrm{H}]_{\mathrm{initial}} < 0.2$~dex).
As clearly seen in the metal-rich SAD, there are no stars with $[\mathrm{M}/\mathrm{H}]_{\mathrm{initial}} > 0.2$~dex older than 10~Gyr.
At the lower metallicities, the minimum in the SAD is less pronounced and located at an age of 4~Gyr.
However, the inflection at 10~Gyr is more pronounced due to a lack of stars with sub-solar metallicities and ages in the range 8 to 10~Gyr.
This results in a stronger peak at ages greater than 10~Gyr.

In order to interpret these age-metallicity features it is necessary to break the sample down into smaller groups of stars.
Therefore, we move on to consider subsamples of stars selected based on their kinematics and Galactic positions.

\subsection{Kinematically selected subsamples} \label{sec:vT_slice}

\begin{figure*}
\centering
\includegraphics[width=\textwidth]{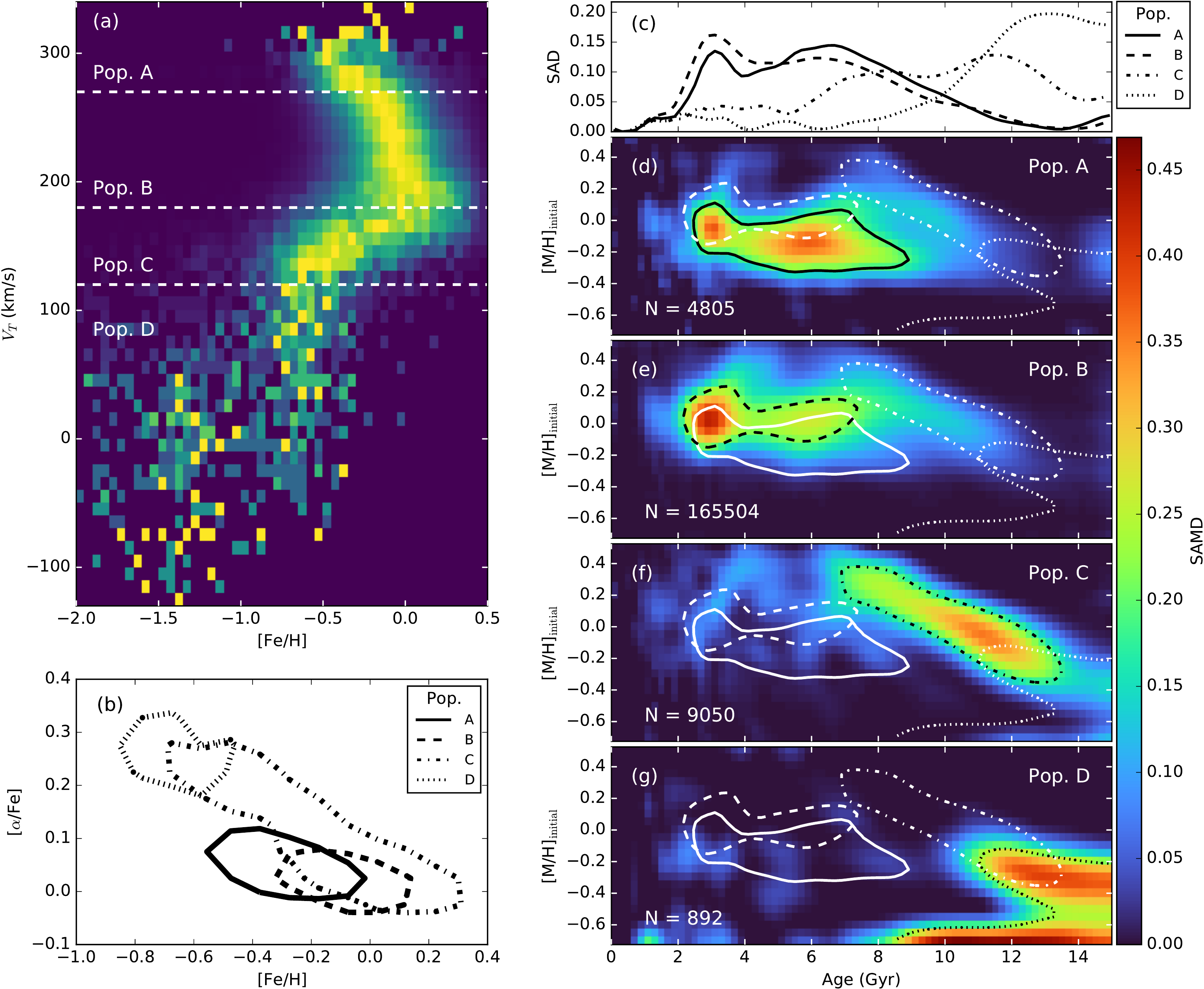}
\caption{The SAMD and SAD in four different bins of tangential velocity, $V_{T}$.
(a) $V_{T}$ against [Fe/H] as a row-normalised 2D histogram.
The dashed lines indicate the division between the four populations, A to D, at $V_{T} = 120$, 180, and 270~km~s$^{-1}$.
(b) Distribution of alpha abundance against metallicity for each population in the form of a half-of-maximum density contour line.
(c) SAD of the four populations.
(d)-(g) SAMD of the four populations.
In each panel the half-of-maximum contour line of the SAMD is shown in black and the contour lines from the other SAMDs are shown in white.}
\label{fig:SAMD_vT_bins}
\end{figure*}

There are many ways to slice a sample kinematically, but in this study we consider a simple division into four populations based on the azimuthal component of the Galactrocentric velocity in cylindrical coordinates, $V_{T}$.
This parameter, which we refer to as the tangential velocity, is taken from the GALAH DR3 VAC.
The division into four populations, A to D, is made by eye in the row-normalised histogram of $V_{T}$ against [Fe/H] as shown in \autoref{fig:SAMD_vT_bins}a.\footnote{In rough terms, Population D is a halo-like population, Population C has the properties of a kinematically defined thick disc, and Populations A and B of a kinematically defined thin disc.}
This approach is inspired by \citet{2020MNRAS.494.3880B} who used this space to identify and study ``the Splash'', an ancient part of the stellar disc thought to have formed \textit{in situ} and been kinematically heated during the last major merger.
The Splash is seen at the highest metallicities of Population D.
These four populations divided according to tangential velocity also show significantly different distributions in the alpha abundance against metallicity plane (\autoref{fig:SAMD_vT_bins}b).

\begin{figure}
\centering
\includegraphics[width=\columnwidth]{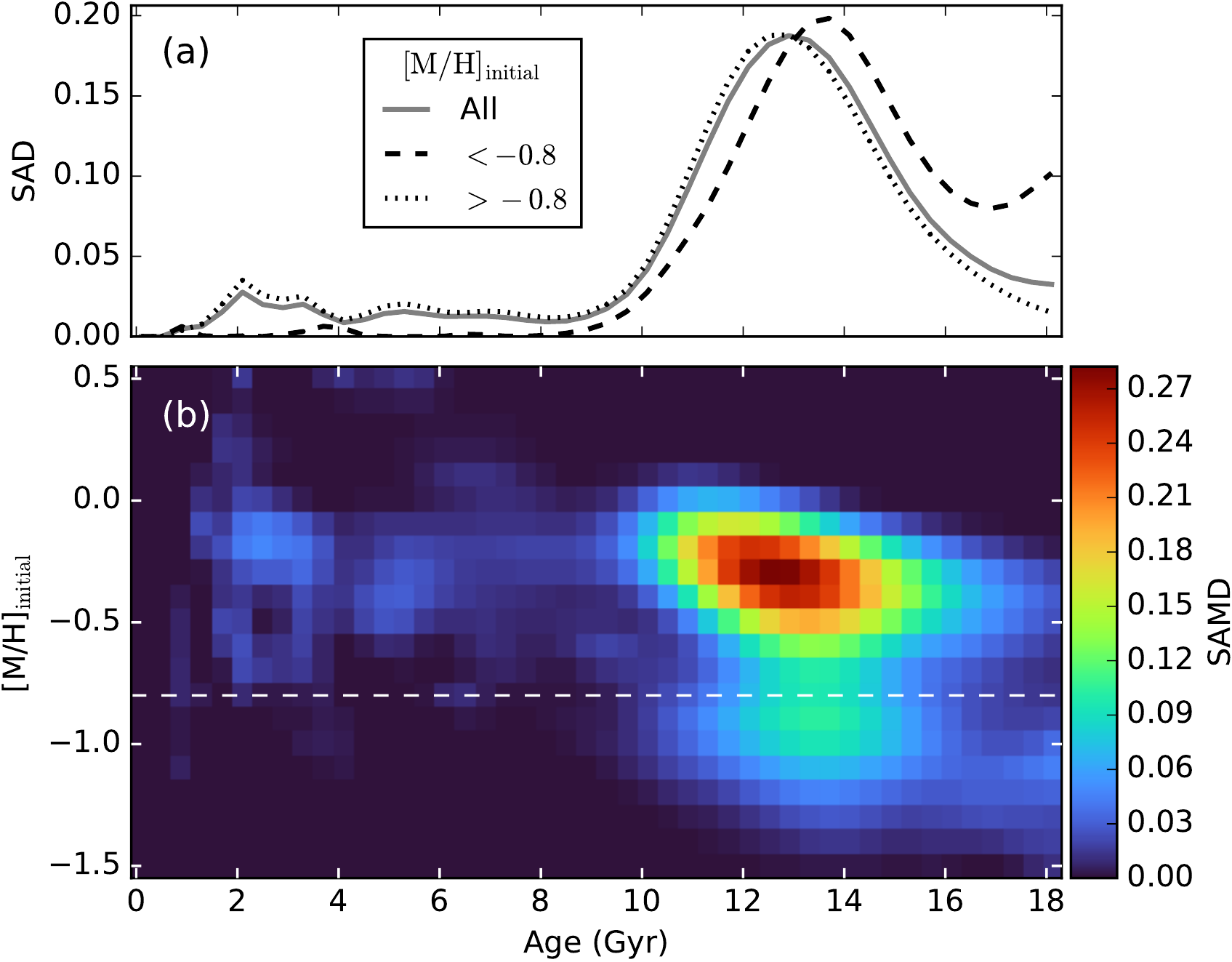}
\caption{The SAD (a) and SAMD (b) of Population D from \autoref{fig:SAMD_vT_bins} ($V_{T} < 120$~km~s$^{-1}$).
The white dashed line across the SAMD indicates $[\mathrm{M}/\mathrm{H}]_{\mathrm{initial}} = -0.8$~dex which is a local minimum in the metallicity distribution.}
\label{fig:SAMD_popD}
\end{figure}

Starting at the lowest velocities, Population D contains the stars with $V_{T} < 120$~km~s$^{-1}$.
\autoref{fig:SAMD_vT_bins}g shows the SAMD of the population which has two low-metallicity peaks between 10 and 15~Gyr.
Since much of Population D falls outside of the main grid of stellar models, the SAMD has been re-calculated using the larger low-resolution grid (\autoref{fig:SAMD_popD}).
The SAMD has a local minimum along the metallicity axis at $[\mathrm{M}/\mathrm{H}]_{\mathrm{initial}} = -0.8$~dex which separates the likely \textit{in situ} population from an accreted population.
As seen in \autoref{fig:SAMD_popD}a, the bulk of these populations are older than 10~Gyr, in agreement with previous studies of the age distribution of halo stars \citep{2019NatAs...3..932G, 2020MNRAS.494.3880B}.
The SAMDs for Populations A to C are all contained within the main model grid, and calculations with the larger low-resolution grid do not change their positions in the age-metallicity plane (see Appendix~\ref{app:lowres_SAMD} and \autoref{fig:SAMD_vT_bins_lowres}).

Population C contains the stars with $V_{T}$ in the range $120$ to $180$~km~s$^{-1}$.
The low-metallicity end of this population connects smoothly with the high-metallicity end of Population D both kinematically and chemically as seen in \autoref{fig:SAMD_vT_bins}a,b.
The SAMD also overlaps with that of Population D at the old metal-poor end and shows a clear age-metallicity relation down to ages around 7~Gyr.
There is a shallow local minimum in the SAD at 10~Gyr but otherwise the age distribution is almost flat between 7 and 12~Gyr.
This population straddles the [$\alpha/$Fe] distribution with the old low-metallicity end at high [$\alpha/$Fe] and the young high-metallicity end at low [$\alpha/$Fe] (\autoref{fig:SAMD_vT_bins}b).

Populations A and B together contain the stars with $V_{T} > 180$~km~s$^{-1}$.
They are divided into Population A with $V_{T} > 270$~km~s$^{-1}$ and Population B with $V_{T} < 270$~km~s$^{-1}$.
Population B contains roughly 92 per cent of the full sample which makes the SAMD in \autoref{fig:SAMD_vT_bins}e look quite similar to the SAMD of the full sample (\autoref{fig:SAMD_all}b).
At the old end, the distribution overlaps with Population C, but the bulk of the distribution is at ages below 8~Gyr with a peak near 3~Gyr.
Population A shows an age distribution similar to Population B but it is slightly older on average and at a lower metallicity.
At the old end it overlaps slightly with Population C at $[\mathrm{M}/\mathrm{H}]_{\mathrm{initial}} \approx 0$~dex, but there is also a lower metallicity tail which does not overlap with any of the other populations.

\subsection{Spatially selected subsamples} \label{sec:Rz_slice}

\begin{figure}
\centering
\includegraphics[width=\columnwidth]{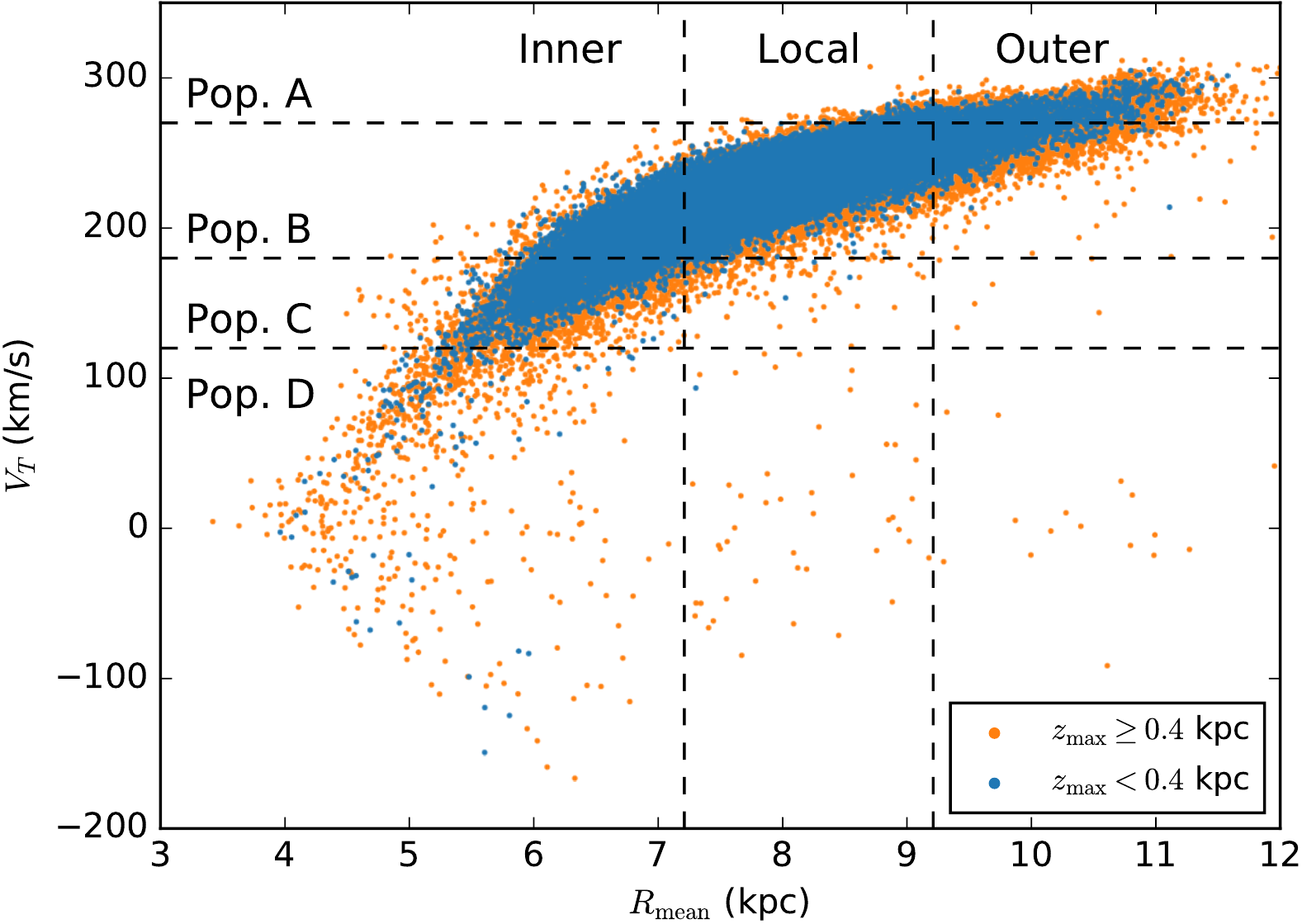}
\caption{Tangential velocity, $V_{T}$, against mean Galactocentric radius, $R_{\mathrm{mean}}$.
The points are coloured according to $z_{\mathrm{max}}$ with orange and blue indicating values above and below $0.4$~kpc, respectively.
The four bins in $V_{T}$ used in \autoref{fig:SAMD_vT_bins} are indicated by horizontal dashed lines, and the three bins in $R_{\mathrm{mean}}$ used in \autoref{fig:SAMD_Rz_bins} are indicated by vertical dashed lines.}
\label{fig:vT_Rmean}
\end{figure}

\begin{figure*}
\centering
\includegraphics[width=\textwidth]{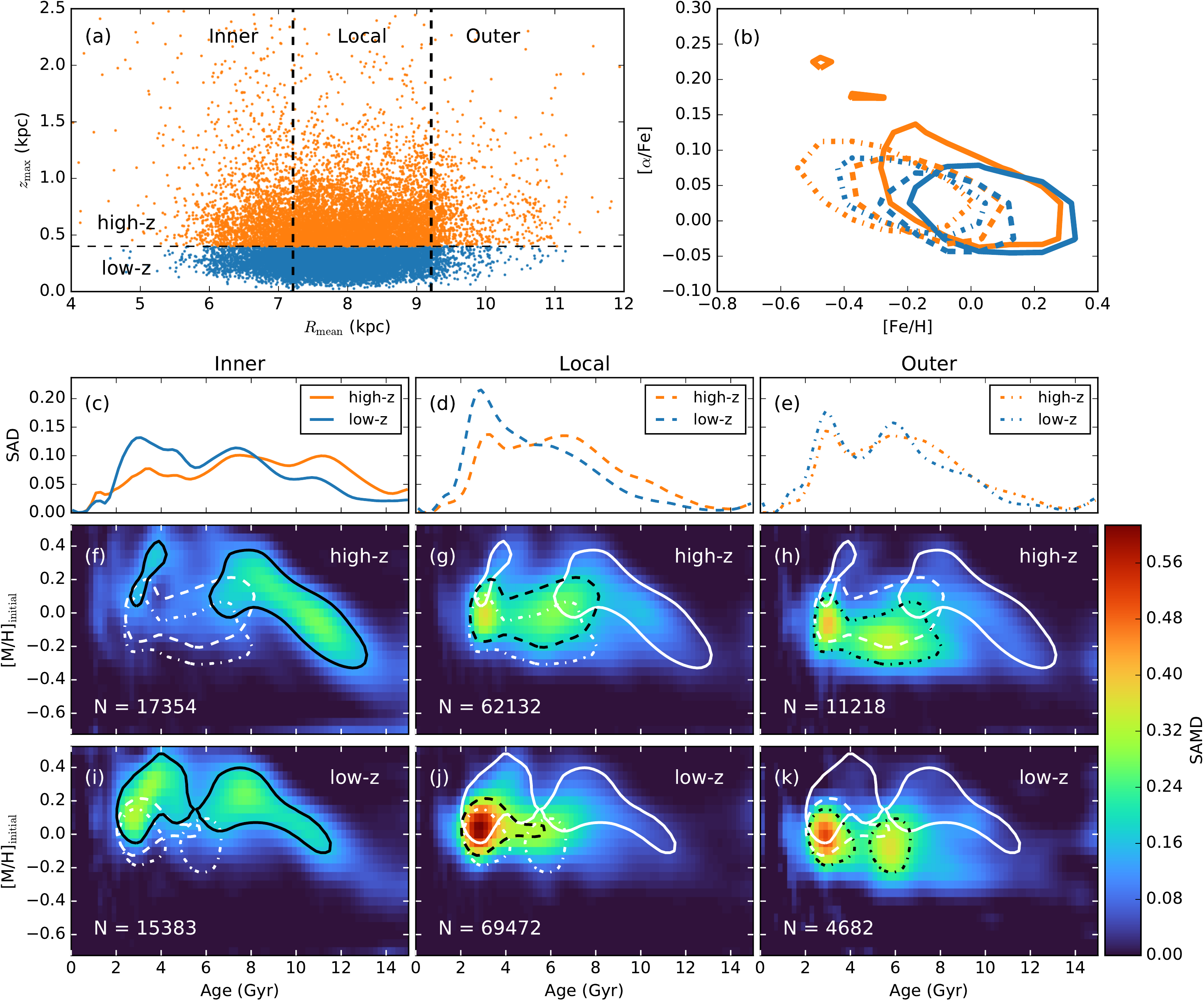}
\caption{The SAMD and SAD in six different bins of $R_{\mathrm{mean}}$ and $z_{\mathrm{max}}$.
(a) $R_{\mathrm{mean}}$ against $z_{\mathrm{max}}$.
The vertical dashed lines indicate the division into three $R_{\mathrm{mean}}$ bins (inner, local, and outer) at $R_{\mathrm{mean}} = 7.21$ and $9.21$~kpc.
The horizontal dashed line indcates the division of each $R_{\mathrm{mean}}$ bin into a high-$z_{\mathrm{max}}$ and low-$z_{\mathrm{max}}$ bin at $z_{\mathrm{max}} = 0.4$~kpc.
(b) Distribution of alpha abundance against metallicity for each population in the form of a half-of-maximum density contour line.
(c)-(e) SAD of the high-$z$ and low-$z$ populations in the inner, local, and outer $R_{\mathrm{mean}}$ bin, respectively.
(f)-(k) SAMD of the six populations.
The panels are arranged according to the division in panel (a).
In each panel the half-of-maximum contour line of the SAMD is shown in black and the contour lines from the other two SAMDs at the same $z_{\mathrm{max}}$ are shown in white.
The line styles and colours in panel (b) can be matched to a spatial bin using the legends in panels (c)-(e).}
\label{fig:SAMD_Rz_bins}
\end{figure*}

The division of the stars into bins of tangential velocity is also to some degree a division into spatial bins.
This is shown in \autoref{fig:vT_Rmean}, which illustrates the correlation between the tangential velocity and the Galactic position in terms of $R_{\mathrm{mean}} = (R_{\mathrm{peri}} + R_{\mathrm{ap}})/2$ and $z_{\mathrm{max}}$, i.e. the mean galactocentric radius and maximum vertical distance from the disc plane.
The use of $R_{\mathrm{mean}}$ and $z_{\mathrm{max}}$ instead of current positions $R$ and $z$ is motivated by two main considerations.
Firstly, $R_{\mathrm{mean}}$ is a better approximation to the birth radius of a star because it averages over the radial oscillations during the orbit, also called blurring.
However, it does not remove the effect of churning which moves stars across the disc without significantly increasing the radial oscillation amplitude \citep{2002MNRAS.336..785S}.
Secondly, it helps reduce selection effects introduced by the magnitude limit of the survey.
If we were to study the age distribution in bins of $R$ and $z$, the most distant populations, which have the largest fraction of intrinsically bright stars, would be among the youngest.

Overall, the tangential velocity is correlated with $R_{\mathrm{mean}}$.
At the lowest velocities (Population D) the majority of stars have $z_{\mathrm{max}} > 0.4$~kpc and small $R_{\mathrm{mean}}$ (\autoref{fig:vT_Rmean}).
Populations A and C contain almost exclusively outer and inner disc stars, respectively.
This can be understood as stars from the inner disc being near the apocentre of their orbits, and therefore at a point of low tangential velocity, and vice versa for stars from the outer disc.
Population B, on the other hand, contains stars from almost the entire range of $R_{\mathrm{mean}}$.

\autoref{fig:SAMD_Rz_bins}f-k shows the SAMD for six bins in $R_{\mathrm{mean}}$ and $z_{\mathrm{max}}$.
We refer to three radial bins delimited at $R_{\mathrm{mean}} = 8.21\pm 1.00$~kpc as the inner, local, and outer region of the disc.
The central value of $8.21$~kpc is the assumed Galactocentric radius of the Sun in the GALAH VAC.
Each of the radial bins is divided into a low-$z_{\mathrm{max}}$ and a high-$z_{\mathrm{max}}$ bin at $z_{\mathrm{max}} = 0.4$~kpc as shown in \autoref{fig:SAMD_Rz_bins}a.

In the inner disc there is a clear difference with $z_{\mathrm{max}}$.
The high-$z_{\mathrm{max}}$ SAMD is similar to Population C while the low-$z_{\mathrm{max}}$ SAMD has a prominent younger peak at high metallicity.
This indicates that the high-metallicity part of the Population B SAMD, both the young and old end, belongs to the inner disc.

The mean value of $[\mathrm{M}/\mathrm{H}]_{\mathrm{initial}}$ at low $z_{\mathrm{max}}$ decreases from about $0.2$~dex in the inner disc to $-0.1$~dex in the outer disc in accordance with the observed negative radial metallicity gradient \citep[e.g.,][]{2014A&A...566A..37G}.
Additionally, as seen in \autoref{fig:SAMD_Rz_bins}, the mean metallicity is lowest at high $z_{\mathrm{max}}$ in each radial bin.

The local and outer regions are younger on average than the inner region with most stars younger than 8~Gyr like Populations A and B.
The SAMD of the local high-$z_{\mathrm{max}}$ population shows two age peaks which connect to the two peaks at higher metallicity in the inner disc at low $z_{\mathrm{max}}$.
These peaks are also present in the outer disc populations indicating that the age distribution is bimodal over the entire range of galactocentric radii probed here, from 5 to 11~kpc.

\section{Discussion} \label{sec:discussion}

\subsection{Impact of the selection function} \label{sec:selection_function}

\begin{figure}
\centering
\includegraphics[width=\columnwidth]{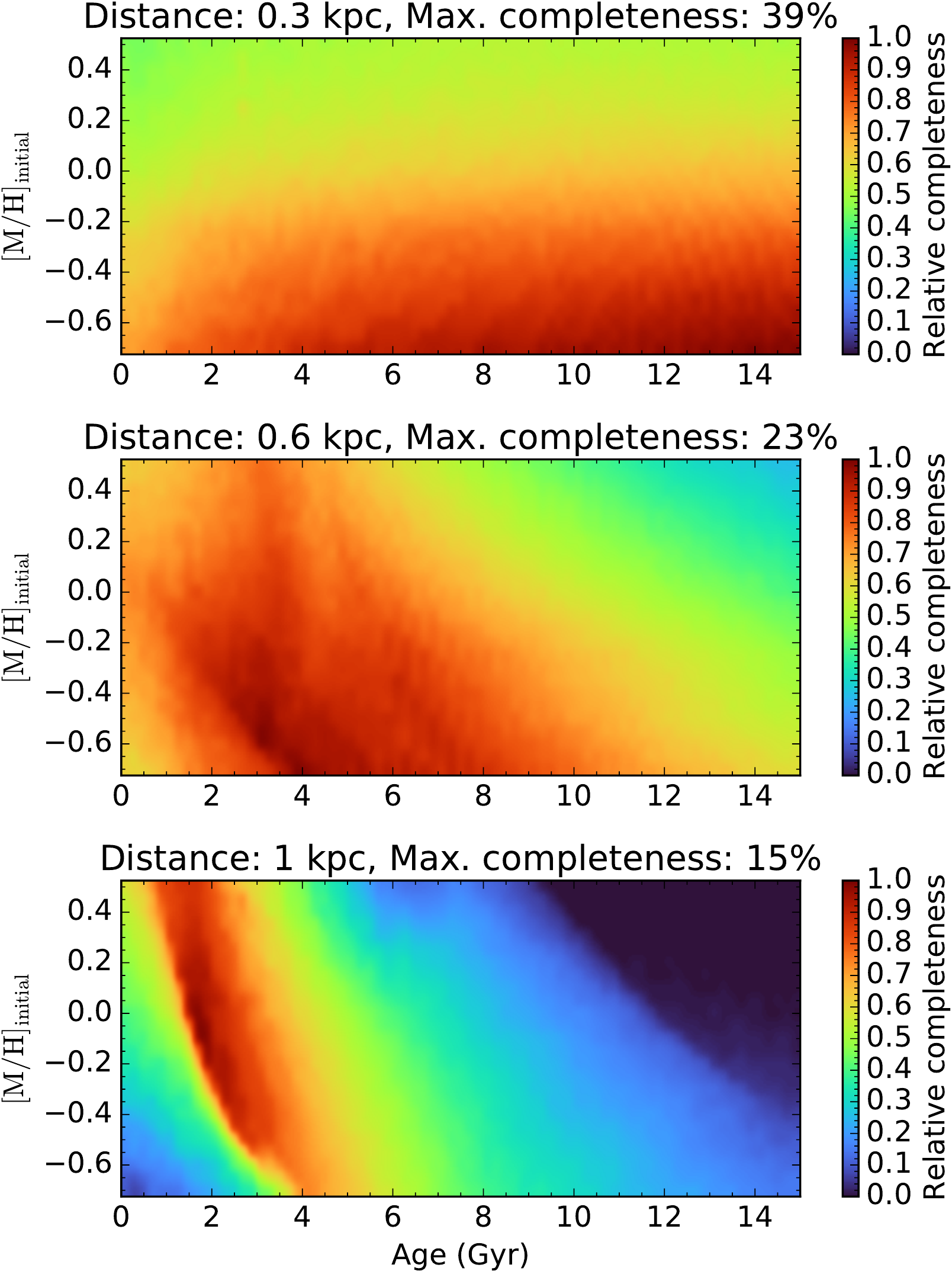}
\caption{The age-metallicity completeness obtained by applying the GALAH selection function (see the text in Section~\ref{sec:selection_function}) to a synthetic sample placed at three different distances.
The colour scale in each panel is normalized relative to the maximum value which is given in the title of the panel.
The mode of the distance distribution of the main GALAH sample is 0.6~kpc and about 85 per cent of the sample is within 1~kpc.}
\label{fig:completeness}
\end{figure}

The age-metallicity distributions presented in the previous section are shaped in part by the selection function of the sample.
The GALAH survey is limited to the magnitude range $12 < m_{V} < 14$ and, although no colour limits are imposed, the survey is effectively limited to the temperature range $4000 < T_{\mathrm{eff}} < 7000$~K \citep{2017MNRAS.465.3203M}.
Since young stars are brighter on average than older stars, the magnitude limit can skew the age distribution towards the young end.
This may partly explain the fact that the age distribution peaks at an age of 3~Gyr.
As shown by \citet{2014A&A...565A..89B}, magnitude and colour limits can act to strongly suppress old metal-rich and young metal-poor stars.

The peak in the age distribution at 3~Gyr is similar to those found in samples of giant stars \citep[e.g.,][]{2016ApJ...817...40F}.
For such samples, the selection function can lead to an excess of young red clump stars compared to the older red giant branch stars, causing the observed age distribution to be bimodal.
This effect, however, cannot explain the young peak found in this work since all giants have been excluded.

To get a better sense of the impact of the GALAH selection function, we performed the test described by \citet{2014A&A...565A..89B}.
This consists of creating a synthetic sample of stars drawn from the PARSEC isochrones with a Salpeter initial mass function, a constant star formation rate, and a flat metallicity distribution.
The synthetic sample is assigned a distance, $d$, in order to calculate apparent magnitudes $m_{V} = M_{V}+5 \log10(d\,\mathrm{[pc]})-5$.
Then a GALAH-like selection is made by taking the subsample with $12 < m_{V} < 14$, $4000 < T_{\mathrm{eff}} < 7000$~K, and by cutting out the giants using the selection $m_{K_{s}} > 8.5-T_{\mathrm{eff}}/(700~\mathrm{K})$ which is equivalent to Eq.~\eqref{eq:magnitude_cut}.
At each age and metallicity the completeness of the selection can be estimated as the number of selected stars divided by the total number.
This completeness function is shown for three distances in \autoref{fig:completeness}.
At a distance of 0.6~kpc, which is the mode of the distance distribution of the main GALAH sample, the completeness is highest at ages around 2 to 4~Gyr.
Unlike the case of a sample of giant stars there is no secondary peak.
The completeness function decreases monotonically on both the young and old side of the peak with some dependence on the metallicity.
Similar monotonic behaviour is seen in the completeness of more nearby and distant stars, although with a greater contrast between the young and old end for the most distant sample.
Overall, the selection function enhances the young peak seen in the SAMD at an age of 3~Gyr but it does not explain the local SAMD minima.

This simple test does not take into account the selection caused by quality cuts or the spatial variations of the true age-metallicity distribution.
For this reason, we do not apply any corrections for sample completeness to the SAMDs; however, an attempt at such a correction is described in Appendix~\ref{app:SAMD_correction}.
Overall, this analysis indicates that magnitude and temperature limits alone are unable to explain the minima in the SAMD seen e.g. in the full sample in \autoref{fig:SAMD_all}.

\subsection{A recent burst of star formation}

One of the persistent features of the SAMD is a prominent young peak at an age of around 3~Gyr.
This peak is seen in all the subsamples containing a significant number of young stars such as the kinematic Populations A and B (\autoref{fig:SAMD_vT_bins}d,e), and the low-$z$ spatial populations (\autoref{fig:SAMD_Rz_bins}i-k).
What makes the peak stand out is the fact that the age distribution has a minimum separating it from a second, older peak at ages around 6 to 8~Gyr.
As discussed in Section~\ref{sec:selection_function}, it is difficult to explain this minimum in terms of the most obvious selection effects.

A similar recent peak in the star formation rate has been found by \citet{2019A&A...624L...1M} based on fitting a Galaxy model to \textit{Gaia} colour-magnitude data, and by \citet{2019ApJ...878L..11I} based on the age distribution of massive white dwarfs.
In both of these studies the star formation rate reaches a maximum 2 to 3~Gyr ago in agreement with the peak we find in the SAMD.
\citet{2021arXiv210309838J} was inspired by these studies to implement a late-burst star formation history in their Galactic chemical evolution model.
They find that this model provides a good fit to the age-metallicity relation of APOGEE stars \citep{2018MNRAS.477.2326F}.
In their model, the burst of star formation is fueled by a late enhancement in the rate of pristine gas accretion.
This gas dilutes the interstellar medium causing most of the stars formed during the burst in the inner Galaxy to have metallicities around the solar value.
Following the burst, the metallicity increases again so inner disc stars are born with super-solar metallicities in the past 2~Gyr.
The effect of this dilution and subsequent metallicity enrichment is a bimodal age distribution at super-solar metallicities.
We do see such a bimodality in the inner disc SAMD (\autoref{fig:SAMD_Rz_bins}i); however, the young peak in the SAMD is at super-solar metallicities already at the time of the burst, 2 to 3 Gyr ago.
This seems to indicate that the burst we find in the inner disc occurs in an already enriched medium.

Another possibly related result is the identification by \citet{2020NatAs...4..965R} of a number of recent peaks in the star formation rate obtained by colour-magnitude diagram fitting of \textit{Gaia} data within about 2~kpc.
They interpret these peaks, located 5.7, 1.9 and 1.0~Gyr ago, as being induced by the pericentre passages of the Sagittarius dwarf galaxy.
The most prominent of the peaks is the one at 5.7~Gyr which is significantly different from the age of 2 to 3~Gyr found in the works mentioned above and in this work.
Since our results are based on a sample covering a similar volume, out to distances of 2~kpc, this is an interesting discrepancy which seems too large to be caused by systematic uncertainties in the age distributions.

\citet{2020A&A...640A..81N} find two sequences in the age-metallicity distribution of solar-type stars with high precision elemental abundances.
Each of the sequences decrease in metallicity with increasing age and they are separated by a lack of stars at an age of 5~Gyr at $[\mathrm{Fe}/\mathrm{H}] = 0$~dex.
As noted by \citet{2020A&A...640A..81N}, a similar lack of stars at intermediate ages was seen by \citet{2015MNRAS.452.2127S} in a sample of stars with asteroseismic ages.
These two studies were both based on samples of fewer than 100 stars within 0.5~kpc of the Sun, but the minimum in the age distribution agrees well with the one we find in this study based on 180\,000 stars.
However, in this study the minimum separates two age-metallicity ridges with older stars at higher metallicities which is opposite to the trend found by \citet{2020A&A...640A..81N}.
This could be due to the extended volume of this study.
We have already shown that the GALAH sample contains both inner disc stars which are old and metal-rich and outer disc stars which are younger and more metal-poor.
When considered together, the age-metallicity trend is the inverse of what one would expect from a chemical evolution sequence where the oldest stars are the most metal-poor.

Altogether, this work adds an additional indication of a real deficit of stars at intermediate ages and a burst of star formation in the past few gigayears.

\subsection{The inner disc age-metallicity relation}

\begin{figure}
\centering
\includegraphics[width=\columnwidth]{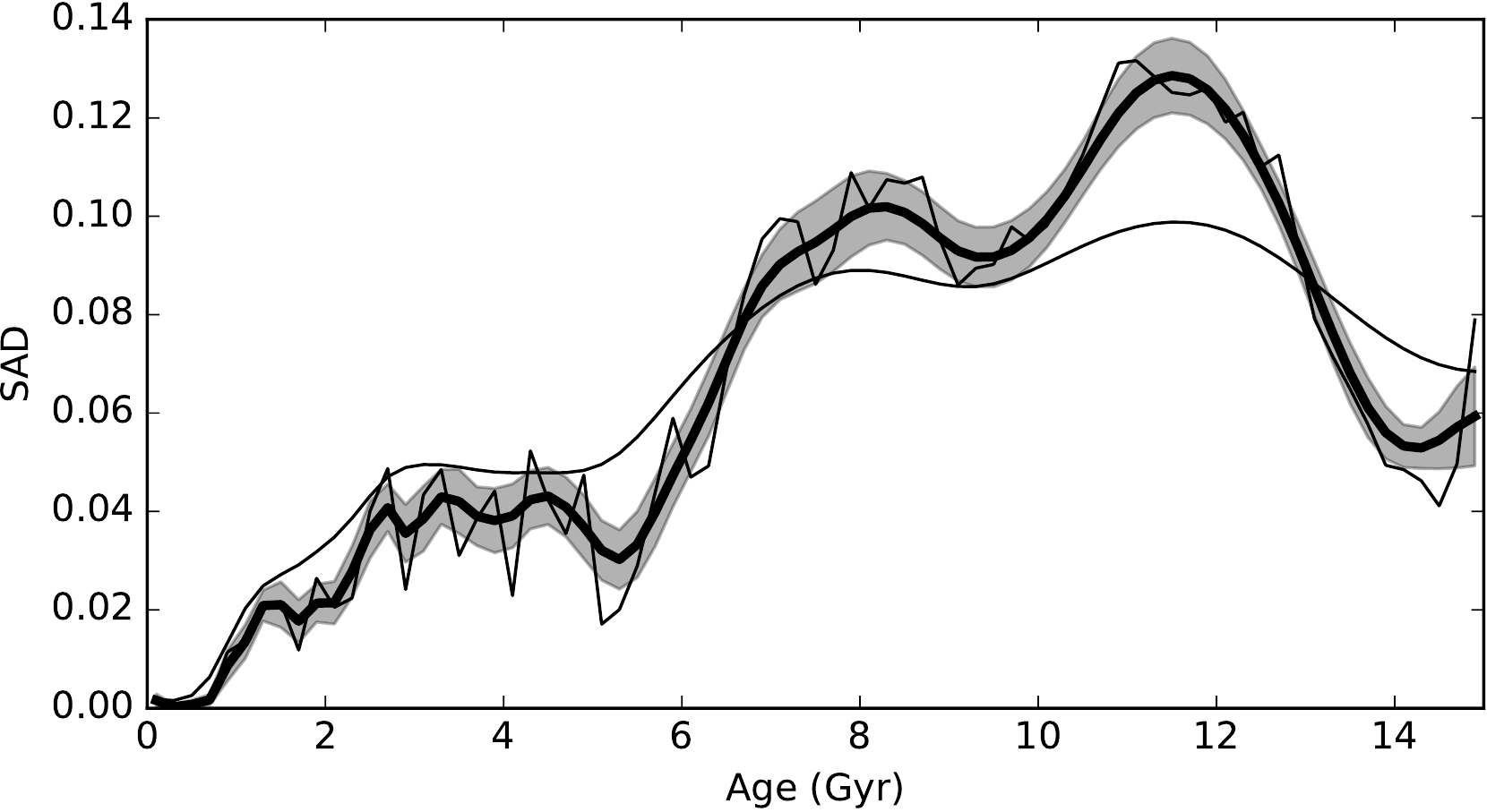}
\caption{SAD of population C.
The thick black line is the same as shown in \autoref{fig:SAMD_vT_bins}c and the grey region indicates the uncertainty calculated by bootstrapping.
The two thin lines show the SAD with no smoothing (a regularization parameter of zero) and with the smoothing increased by a factor of 30 compared to the fiducial value.
}
\label{fig:popC_SAD}
\end{figure}

\begin{figure}
\centering
\includegraphics[width=\columnwidth]{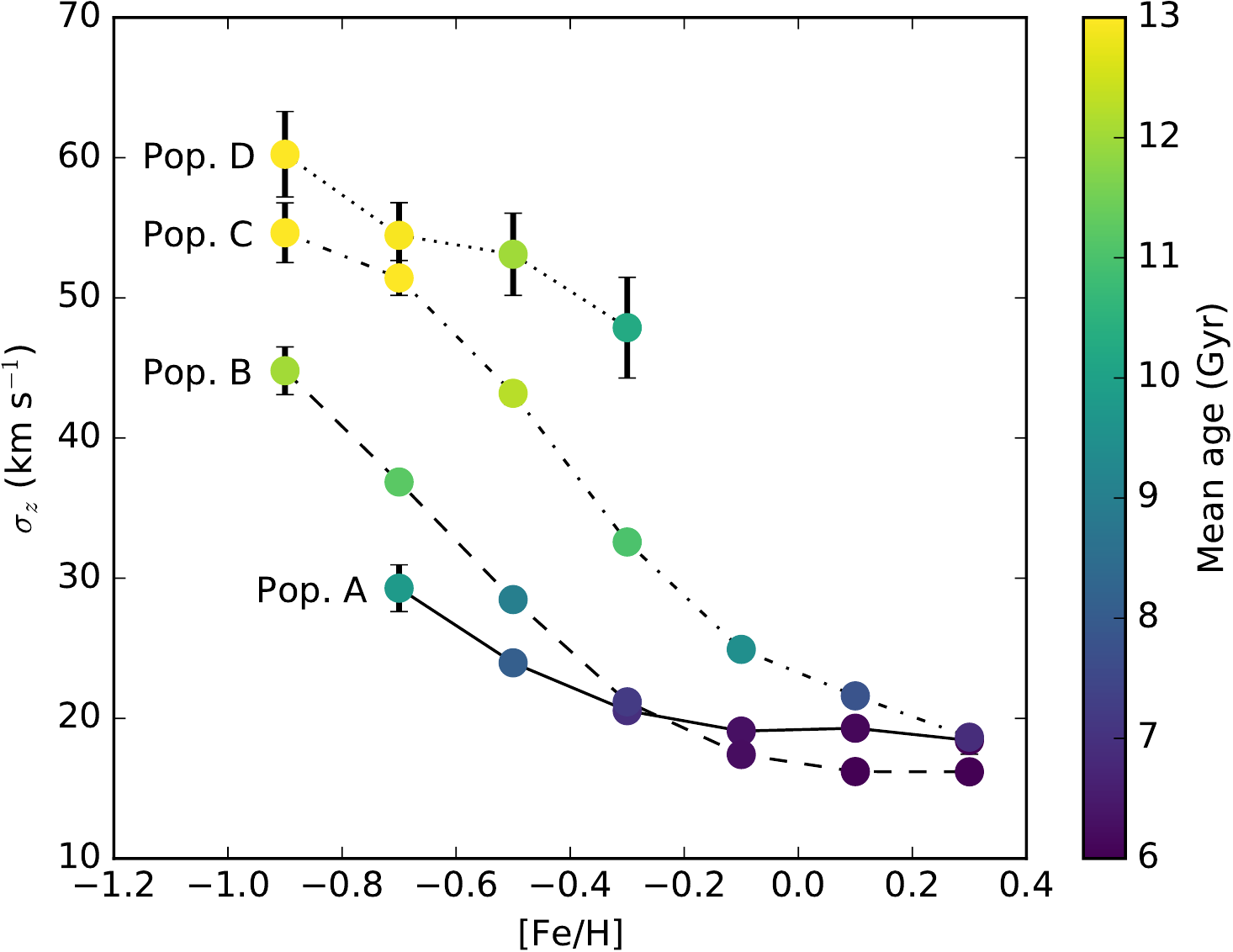}
\caption{Vertical velocity dispersion binned by metallicity for the four kinematic populations.
Errorbars have been estimated using bootstrapping.
The bin width is $0.2$~dex and the points are plotted in the middle of each bin.
The colour indicates the mean of the individual age estimates of the stars in each bin.}
\label{fig:vz_disp_feh}
\end{figure}

Population C shows a clear continuous sequence in the age-metallicity distribution from old and metal-poor to intermediate ages and metal-rich (\autoref{fig:SAMD_vT_bins}f).
This age-metallicity sequence belongs mainly to the inner disc and is seen most clearly for stars with high $z_{\mathrm{max}}$ (\autoref{fig:SAMD_Rz_bins}f).
A similar age-metallicity relation in the inner disc has been found previously and associated with the thick disc as defined by the stellar kinematics \citep{2004A&A...421..969B}.
It has also been found in the chemically defined thick disc, i.e. stars with high alpha abundances, and interpreted as the early chemical evolution in a well-mixed inner Galaxy \citep{2013A&A...560A.109H}.
Thus, the age-metallicity relation of Population C could be interpreted as the thick disc chemical evolution sequence.

As a counterargument to calling Population C the thick disc, we note that it includes a significant amount of stars with low alpha abundances at the high-metallicity, young end (see \autoref{fig:SAMD_vT_bins}b).
There is a local minimum in the age distribution at $10$~Gyr, as shown in \autoref{fig:popC_SAD}, which seems to mark the transition from high to low alpha abundance with increasing metallicity.
We have checked that this local minimum in the age distribution is robust against the level of smoothing used when calculating the SAMD.
The two thin lines in \autoref{fig:popC_SAD} show the SAD of Population C for a lower and higher value of the regularization parameter compared to the fiducial value.
We find that the presence and locations of the main peaks and minima are largely unaffected by the degree of smoothing.


Population C also shows a sharp transition in vertical velocity dispersion, $\sigma_{z}$, from $55$~km~s$^{-1}$ at the metal-poor end to $20$~km~s$^{-1}$ at the metal-rich end (\autoref{fig:vz_disp_feh}).
The velocity dispersion is correlated with age such that the stars with $\sigma_{z} \gtrsim 30$~km~s$^{-1}$ have a mean age above 10~Gyr and vice versa.
The extreme values of $55$ and $20$~km~s$^{-1}$ are consistent with the values usually associated with the thick and thin discs, respectively \citep{2016ARA&A..54..529B}.

This distinction between the low- and high-metallicity inner disc populations is similar to the results of other works.
For example, \citet{2017A&A...608L...1H} studied about 500 nearby stars and found that only the most metal-poor stars of the high-alpha sequence are vertically extended.
More metal-rich stars in the high-alpha sequence were found to have vertical velocity dispersions around the level of, or slightly lower than, metal-poor stars in the low-alpha sequence.
Additionally, \citet{2018A&A...619A.125A} applied the dimensionality reduction algorithm t-SNE on the space of 13 different elemental abundances for a sample of about 1000 local stars and found that the high-alpha stars separate into a number of distinct clusters.
The metal-poor end, which they label thick disc, separates from the rest of the high-alpha sequence.
These results imply a transition from a thick to thin inner disc -- as defined in terms of kinematics and abundances -- along a single chemical evolution sequence with increasing [Fe/H].
The SAD of population C indicates that this transition occurred roughly 10~Gyr ago and that it was accompanied by a dip in the star formation rate.


\subsection{Formation of the stellar disc -- our results compared to different scenarios and models}

\begin{figure*}
\centering
\includegraphics[width=\textwidth]{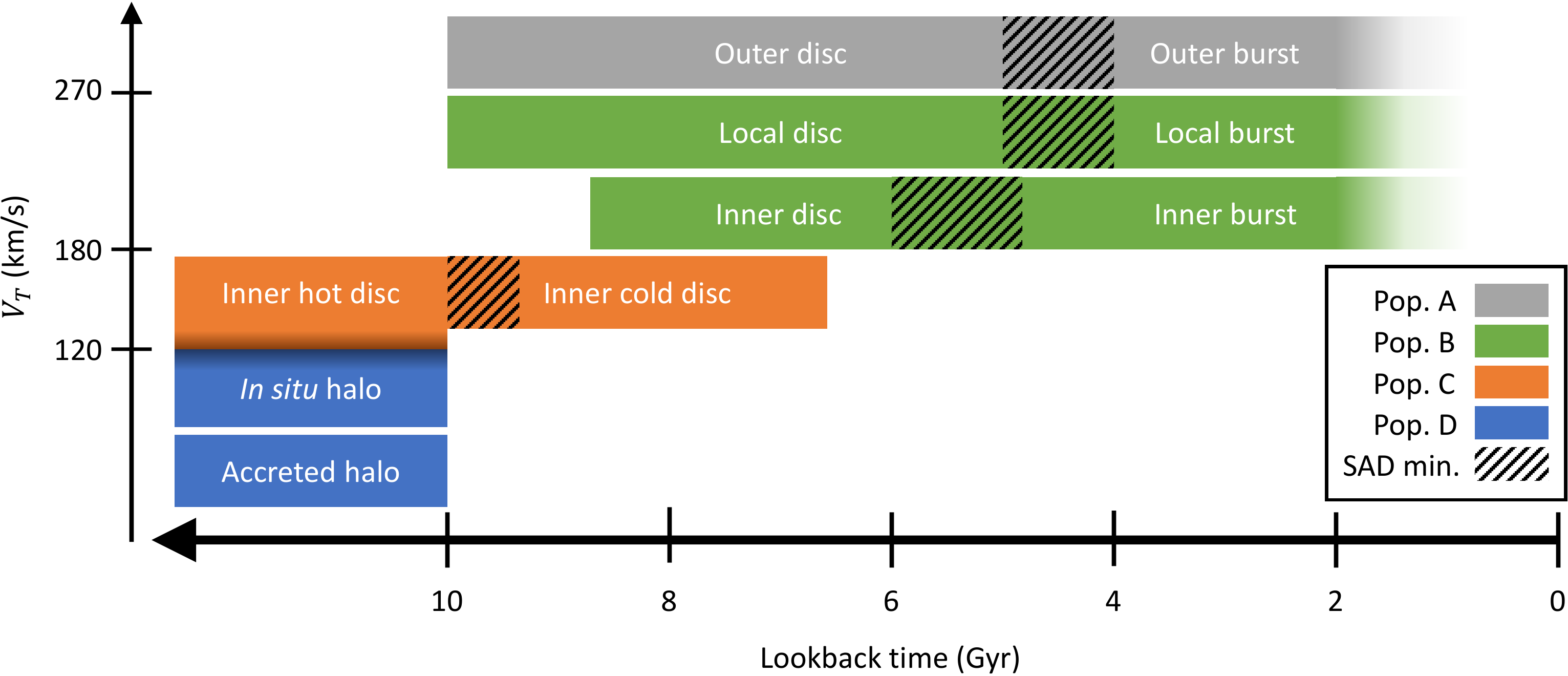}
\caption{Schematic timeline giving an overview of the age distributions of the four kinematically selected subsamples shown in \autoref{fig:SAMD_vT_bins}.
Populations C and D are connected at the old end to indicate that one transitions smoothly into the other in the inner disc.
The hatched regions mark the approximate ages at which minima are found in the SAD.}
\label{fig:timeline}
\end{figure*}

We summarise the results discussed so far in terms of a timeline shown in \autoref{fig:timeline}.
The timeline is a schematic representation of when the different populations of the GALAH sample formed and where the minima in the age distributions occur.
The first transition is found to be around 10~Gyr ago.
Stars on the old side of this transition belong either to an accreted population or the inner disc.
Disc stars at these ages are kinematically hot with vertical velocity dispersions greater than $30$~km~s$^{-1}$.
There is a smooth transition in age, metallicity, alpha abundance, and $V_{T}$ between the \textit{in situ} halo in Population D and the inner hot disc of Population C.
Stars younger than the 10~Gyr transition are found in all regions of the Galaxy covered by the GALAH sample.
In the inner disc, the metallicity continues to increase with decreasing age and the velocity dispersion drops below $30$~km~s$^{-1}$ motivating the label ``inner cold disc'' in \autoref{fig:timeline}.
The youngest inner disc stars are also found at high metallicities and there is a minimum in the SAD at 5~Gyr separating them from the older stars.
Both local and outer disc stars seem to have begun forming shortly after the 10~Gyr transition but at different metallicities.
Especially the oldest outer disc stars in Population A stand out with many of them having a metallicity about 0.5~dex lower than the inner disc stars at the same age (see \autoref{fig:SAMD_vT_bins}d).
Both the local and outer disc populations show a minimum in the SAD about 4~Gyr ago separating the early disc formation from a recent burst.

Altogether the results indicate that the disc has formed in three main phases.
Stars formed in the first phase are found on kinematically hot orbits in the inner disc.
In the second phase, stars are found on kinematically cool orbits across the entire disc with a negative radial metallicity gradient.
Finally, a recent burst of star formation marks the third phase.
This burst seems to occur slightly earlier in the inner disc compared to the local and outer disc.

The idea that the stellar disc has formed in separate phases is reminiscent of the two-infall model \citep{1997ApJ...477..765C}.
Our results indicate that the second infall in such a model must happen around the 10~Gyr transition at the time when the local and outer disc begin to form at a lower metallicity than the inner disc.
This is roughly the time found for the second infall in the revised two-infall model by \citet{2019A&A...623A..60S} which is designed to reproduce the two alpha-abundance sequences observed in the solar neighborhood.
However, at this point in time we find that the low-metallicity stars in the outer disc begin to form while the inner disc is still increasing in metallicity.
This does not fit into the picture of a one-zone two-infall model in which the interstellar medium is well mixed and therefore the metal-poor low-alpha stars only begin to form after the highest metallicities have been reached in the first infall episode.
Perhaps a multi-zone two-infall model as explored by \citet{2021A&A...647A..73S} can reproduce the simultaneous formation of metal-rich stars in the inner disc and metal-poor stars in the outer disc.

Our timeline is overall very similar to the Milky Way disc formation scenario outlined by \citet{2019A&A...625A.105H}.
In their scenario, an inner stellar disc forms first out of a well-mixed turbulent medium with initially low metallicity and high alpha abundance.
After this phase, star formation quenches between 7 and 9~Gyr ago and at the edge of the inner disc the enriched gas left over mixes with more metal-poor gas from the outer disc.
When star formation resumes, the metallicity in the inner disc continues to rise to super-solar values while the outer disc builds up at lower metallicities, fueled by external gas inflow.
The idea of a quenching phase separating two phases of star formation in the inner disc fits well with our results, but we find the transition to happen slightly earlier; that is, at ages around 10~Gyr instead of 7 to 9~Gyr.
If these two times refer to the same transition in the disc, the discrepancy may be caused by the different timing methods.
In their scenario, the quenching time is informed by a chemical evolution model made to reproduce the distribution of [Si/Fe] as a function of age for stars in the solar neighborhood \citep{2015A&A...578A..87S, 2016A&A...589A..66H}.
Our result is inferred directly from the age distribution of stars belonging to the inner disc and is therefore independent of any chemical evolution model.
On the other hand, there is always the possibility of systematic uncertainties in the age estimates related to inaccuracies in the stellar models.

A similar scenario is described by \citet{2021MNRAS.503.2814C} who studied a sample of nearly 20\,000 APOGEE giants with ages determined from a machine learning model trained on \textit{Kepler} seismic data.
Their results suggest that a kinematically hot inner disc with low [Fe/H] and high alpha-element abundances forms first from well-mixed and turbulent gas.
After this there is a smooth transition to the high-[Fe/H] low-alpha population with cooler kinematics in the inner disc, similar to what we find in Population C.
Star formation in the outer disc begins after the initial inner disc phase giving rise to the low-alpha population in an inside-out manner with a metallicity gradient.

The scenarios described by \citet{2019A&A...625A.105H} and \citet{2021MNRAS.503.2814C} do not include a late burst of star formation like the one we find in the SAMD.
The lack of this feature in the relatively large sample considered by \citet{2021MNRAS.503.2814C} may be caused by age uncertainties.
Even though their sample is limited to stars with uncertainties on the logarithm of the age of less than 10 per cent (corresponding to a relative age uncertainty of 20 per cent at 6~Gyr), this uncertainty only accounts for the scatter introduced by the machine learning model and not the uncertainties on the training sample.
Therefore, their age distributions are likely less precise than the SADs considered in this work.

The timeline presented here can also be compared with the results of hydrodynamical galaxy simulations.
Many such simulations of galaxies, either in isolation or in a cosmological context, have succeeded in producing distinct high- and low-alpha populations similar to the ones seen in the Milky Way \citep[e.g.,][]{2018MNRAS.474.3629G, 2020MNRAS.491.5435B, 2021MNRAS.501.5176K, 2021MNRAS.503.5826A}.
A common feature of these simulations is an early central burst of star formation creating the bulk of the high-alpha stars.
This leads to a rapid increase in metallicity during the first few gigayears similar to the age-metallicity relation we find in Population C.
Formation of the younger low-alpha disc component is usually associated with a slower and more extended star formation fuelled by accreting gas.
The details of these later stages differ between the simulations.

In the simulations of isolated disc galaxies by \citet{2021MNRAS.501.5176K}, the low-alpha sequence starts forming out of gas accreting from the halo.
Part of this gas has been ejected in the initial period of intense star formation before reaccreting on a longer timescale.
Using cosmological simulations from the NIHAO-UHD project, \citet{2020MNRAS.491.5435B} find similarly that the low-alpha sequence forms out of accreted gas.
However, in the NIHAO simluations the gas is brought in by a gas-rich merger causing a rapid decrese of the ISM metallicity.

In the VINTERGATAN cosmological simulation \citep{2021MNRAS.503.5826A, 2021MNRAS.503.5846R, 2021MNRAS.503.5868R}, the formation of the low-alpha sequence is also associated with the time of a merger.
In this simulation, the merging galaxy does not dilute the existing gas; instead, an extended outer disc, which is initially misaligned with the inner disc (almost orthogonal to it), forms stars from the inflow of metal-poor gas starting around the time of the last major merger. 
In this way, the inner and outer discs initially evolve almost independently leading to the simultaneous formation of stars at a wide range of metallicities.
As a consequence, one of the predictions of the VINTERGATAN simulation is a minimum in the metallicity distribution at ages corresponding to the beginning of the outer disc formation, i.e. around the time of the last major merger.
In the GALAH data, although there is a significant difference between the metallicities of Populations A and C at ages of about $8$~Gyr, stars of intermediate metallicities do exist in the local disc population leaving no clear bimodality in the full sample.
There is perhaps a small hint of a metallicity bimodality at the old end of the outer disc population (\autoref{fig:SAMD_Rz_bins}h,k), but this result is not conclusive.
In the VINTERGATAN scenario, it is possible that the lack of an observed minimum in the metallicity distribution indicates that the inner and outer discs were initially more aligned.
This would lead to earlier mixing of material between the two regions of the disc and an earlier production of stars at intermediate metallicities.

\citet{2019ApJ...883L...5B} investigated a Milky Way analogue in the EAGLE suite of cosmological simulations.
Their focus was on a galaxy with a major merger occurring at a lookback time of 9 Gyr.
The effect of such a merger in this particular simulation was to contribute to the formation of the thick disc through dynamical heating of the existing stars and to significantly increase the star formation rate in the thin disc component.
Such an increase in the star formation rate could potentially explain the young peak we find in the SAD of population C after the minimum at 10 Gyr.
To take this analysis further we would need to be able to compare the age-metallicity distribution of stellar populations in the simulation with the SAMD of Population C.
As remarked by \citet{2019ApJ...883L...5B}, the low resolution of the simulation prevents such a detailed investigation of the properties of the simulated stellar populations.

Although these simulations differ in their details and exact timelines, they suggest that the transition we find at 10 Gyr marks the end of a central starburst during which most of the high-alpha stars are formed.
The subsequent increase in star formation in a kinematically cooler and more extended disc component may be induced by accretion of gas previously ejected or brought in by, or in connection with, a major merger.

The merger scenario is consistent with the interpretation of the \textit{in situ} part of Population D as being the Splash introduced in Section~\ref{sec:vT_slice}.
The Splash as defined by \citet{2020MNRAS.494.3880B} is seen at [Fe/H]~$\gtrsim -0.7$~dex in \autoref{fig:SAMD_vT_bins}a where it connects to Population C.
In every parameter considered here, the Splash is consistent with being the smooth extension of the old end ($> 10$~Gyr) of Population C.
Population D also contains a small number of stars with [Fe/H]~$< -1$~dex and tangential velocities close to zero which were likely accreted as part of the \textit{Gaia}-Sausage-Enceladus \citep[GSE;][]{2018MNRAS.478..611B, 2018Natur.563...85H} or other recently identified halo substructures \citep{2019MNRAS.488.1235M, 2020ApJ...901...48N}.
The GALAH survey is mainly focused on the disc, so this component only contributes a few hundred stars to the total sample.
Both the Splash and GSE populations are known to be made up of stars with ages greater than about $10$~Gyr \citep{2019ApJ...881L..10S, 2019NatAs...3..932G, 2020ApJ...897L..18B}.
Based on the SAMD of Population D (\autoref{fig:SAMD_popD}), the Splash population is slightly younger than the accreted population, by about 1~Gyr on average.
This is consistent with the results of \citet{2020MNRAS.494.3880B}.

As an alternative to scenarios involving mergers, \citet{2020ApJ...891L..30A} has shown that a Splash-like population can develop in an isolated galaxy without a merger event.
They used a hydrodynamical simulation presented by \citet{2019MNRAS.484.3476C} in which stars are scattered to high-velocity orbits during an early clumpy star formation phase.
In this simulation, the high-alpha population is formed in clumps with high star formation rates, primarily in the inner disc.
This mode of star formation dominates in the first few gigayears, after which most stars form in a smooth configuration distributed throughout the disc at a lower star formation rate.
In this interpretation, the transition we find at 10 Gyr does not indicate the renewal of star formation induced by gas accretion or a major merger.
Instead, it would mark the end of the clumpy formation phase when the more distributed star formation became dominant.
It is not clear, however, how the young end of Population C, which forms at high metallicity immediately following the transition, fits into this scenario.
There would have to be a continued smooth star formation from enriched gas in the inner disc following the end of the clump-dominated phase.

The idea of a Splash without a merger is not necessarily in tension with the observation of accreted GSE stars with ages greater than 10~Gyr.
\citet{2020ApJ...902..119D} argue that coherent shell structures observed in the Milky Way halo indicate that a significant merger, with kinematics similar to the GSE, occurred no more than 5~Gyr ago.
They speculate that this debris is from the GSE merger, which collided with the Milky Way only recently after a prolonged orbit following its initial infall about 10~Gyr ago.
One may wonder, in such a scenario, if interactions between the Milky Way and GSE have played a role in inducing star formation following both the main minima we find in the age distributions of disc stars.

As indicated by the preceding discussion, several scenarios can explain the overall trends found in the observed age-metallicity distribution.
One of the challenges for the future is to determine the relative impact on the early star formation history of gas-rich mergers and secular processes, like an early clumpy phase.
To this end, more detailed comparisons between models and observations may provide important insight.
It would be useful, for example, to compare observations with specific quantitative model predictions for the evolution of the vertical velocity dispersion or the age distributions of stars born in different regions of the disc.

\section{Conclusions} \label{sec:conclusions}

In this work we have presented a detailed investigation of the age-metallicity distribution of stars in the Milky Way disc using a sample of $180\,000$ dwarfs and subgiants from the combination of GALAH DR3 and \textit{Gaia} EDR3.
The results are based on a recently developed method for combining the joint age-metallicity probability density functions of inidividual stars into a sample age-metallicity distribution (SAMD).
The SAMD of the full sample shows more structure than usually seen in scatter plots of individual ages and metallicites; most prominently, there is a clear age-metallicity relation for stars older than about 8~Gyr and a minimum in the age distribution between 4 and 6~Gyr.

By breaking the sample down by kinematics and Galactic position ($R_{\mathrm{mean}}$ and $z_{\mathrm{max}}$), we show that the old age-metallicity relation is most prominent in stars with low tangential velocities belonging to the inner disc ($R_{\mathrm{mean}} \lesssim 7$~kpc).
Along this inner disc age-metallicity relation, there is a minimum in the age distribution at 10~Gyr separating kinematically hot, alpha-rich stars from kinematically cooler, alpha-poor stars.
The local and outer disc populations are younger than 10~Gyr and thus start forming after the inner disc transition and at lower metallicites.
At an age of 8~Gyr, the stars with the highest tangential velocities from the outer disc ($R_{\mathrm{mean}} \gtrsim 9$~kpc) are forming at a metallicity about 0.5~dex lower than inner disc stars.
These main results are in good agreement with other recent observational work arguing for distinct formation phases of the inner and outer disc populations.

We have summarised the results in the form of a schematic timeline (\autoref{fig:timeline}) showing the formation periods of the different subsamples.
Overall, the subsamples have formed in three main phases separated by two local minima in the age distributions.
The first minimum, at 10~Gyr, aligns with the proposed time for the GSE merger which may have played a role in the inner disc transition by heating existing disc stars and inducing star formation.
However, we cannot exclude an alternative scenario like the transition from central clumpy to extended smooth star formation.
The most recent phase, at ages between $2$ and $4$~Gyr, aligns with previous claims of a recent burst of star formation in the disc.
This burst is found in all populations investigated here, spanning $R_{\mathrm{mean}}$ of $5$ to $11$~kpc.

In the future it would be interesting to make more detailed comparisons between these observational constraints on the age-metallicity distribution of disc stars and the results of galaxy simulations.
Additionally, the method used here to calculate the SAMD could be adapted to derive [$\alpha$/Fe]-age distributions for subpopulations in the disc.

\section*{Acknowledgements}
CLS, SF, and DF was supported by the grant 2016-03412 from the Swedish Research Council.

This work made use of the Third Data Release of the GALAH Survey (Buder et al. 2021). The GALAH Survey is based on data acquired through the Australian Astronomical Observatory, under programs: A/2013B/13 (The GALAH pilot survey); A/2014A/25, A/2015A/19, A2017A/18 (The GALAH survey phase 1); A2018A/18 (Open clusters with HERMES); A2019A/1 (Hierarchical star formation in Ori OB1); A2019A/15 (The GALAH survey phase 2); A/2015B/19, A/2016A/22, A/2016B/10, A/2017B/16, A/2018B/15 (The HERMES-TESS program); and A/2015A/3, A/2015B/1, A/2015B/19, A/2016A/22, A/2016B/12, A/2017A/14 (The HERMES K2-follow-up program). We acknowledge the traditional owners of the land on which the AAT stands, the Gamilaraay people, and pay our respects to elders past and present. This paper includes data that has been provided by AAO Data Central (datacentral.aao.gov.au).

This work has made use of data from the European Space Agency (ESA) mission
{\it Gaia} (\url{https://www.cosmos.esa.int/gaia}), processed by the {\it Gaia}
Data Processing and Analysis Consortium (DPAC,
\url{https://www.cosmos.esa.int/web/gaia/dpac/consortium}). Funding for the DPAC
has been provided by national institutions, in particular the institutions
participating in the {\it Gaia} Multilateral Agreement.

\section*{Data Availability}

The data underlying this article were accessed from GALAH Data Release 3 (\url{https://docs.datacentral.org.au/galah/dr3/}) and \textit{Gaia} Early Data Release 3 (\url{https://gea.esac.esa.int/archive/}).
The derived data generated in this research will be shared on reasonable request to the corresponding author.



\bibliographystyle{mnras}
\bibliography{GALAH_AMR} 

\begin{thebibliography}{}
\makeatletter
\relax
\def\mn@urlcharsother{\let\do\@makeother \do\$\do\&\do\#\do\^\do\_\do\%\do\~}
\def\mn@doi{\begingroup\mn@urlcharsother \@ifnextchar [ {\mn@doi@}
  {\mn@doi@[]}}
\def\mn@doi@[#1]#2{\def\@tempa{#1}\ifx\@tempa\@empty \href
  {http://dx.doi.org/#2} {doi:#2}\else \href {http://dx.doi.org/#2} {#1}\fi
  \endgroup}
\def\mn@eprint#1#2{\mn@eprint@#1:#2::\@nil}
\def\mn@eprint@arXiv#1{\href {http://arxiv.org/abs/#1} {{\tt arXiv:#1}}}
\def\mn@eprint@dblp#1{\href {http://dblp.uni-trier.de/rec/bibtex/#1.xml}
  {dblp:#1}}
\def\mn@eprint@#1:#2:#3:#4\@nil{\def\@tempa {#1}\def\@tempb {#2}\def\@tempc
  {#3}\ifx \@tempc \@empty \let \@tempc \@tempb \let \@tempb \@tempa \fi \ifx
  \@tempb \@empty \def\@tempb {arXiv}\fi \@ifundefined
  {mn@eprint@\@tempb}{\@tempb:\@tempc}{\expandafter \expandafter \csname
  mn@eprint@\@tempb\endcsname \expandafter{\@tempc}}}

\bibitem[\protect\citeauthoryear{{Agertz} et~al.,}{{Agertz}
  et~al.}{2021}]{2021MNRAS.503.5826A}
{Agertz} O.,  et~al., 2021, \mn@doi [\mnras] {10.1093/mnras/stab322}, \href
  {https://ui.adsabs.harvard.edu/abs/2021MNRAS.503.5826A} {503, 5826}

\bibitem[\protect\citeauthoryear{{Amarante}, {Beraldo e Silva}, {Debattista}
  \& {Smith}}{{Amarante} et~al.}{2020}]{2020ApJ...891L..30A}
{Amarante} J. A.~S.,  {Beraldo e Silva} L.,  {Debattista} V.~P.,   {Smith}
  M.~C.,  2020, \mn@doi [\apjl] {10.3847/2041-8213/ab78a4}, \href
  {https://ui.adsabs.harvard.edu/abs/2020ApJ...891L..30A} {891, L30}

\bibitem[\protect\citeauthoryear{{Amarsi} et~al.,}{{Amarsi}
  et~al.}{2020}]{2020A&A...642A..62A}
{Amarsi} A.~M.,  et~al., 2020, \mn@doi [\aap] {10.1051/0004-6361/202038650},
  \href {https://ui.adsabs.harvard.edu/abs/2020A&A...642A..62A} {642, A62}

\bibitem[\protect\citeauthoryear{{Anders}, {Chiappini}, {Santiago},
  {Matijevi{\v{c}}}, {Queiroz}, {Steinmetz}  \& {Guiglion}}{{Anders}
  et~al.}{2018}]{2018A&A...619A.125A}
{Anders} F.,  {Chiappini} C.,  {Santiago} B.~X.,  {Matijevi{\v{c}}} G.,
  {Queiroz} A.~B.,  {Steinmetz} M.,   {Guiglion} G.,  2018, \mn@doi [\aap]
  {10.1051/0004-6361/201833099}, \href
  {https://ui.adsabs.harvard.edu/abs/2018A&A...619A.125A} {619, A125}

\bibitem[\protect\citeauthoryear{{Belokurov}, {Erkal}, {Evans}, {Koposov}  \&
  {Deason}}{{Belokurov} et~al.}{2018}]{2018MNRAS.478..611B}
{Belokurov} V.,  {Erkal} D.,  {Evans} N.~W.,  {Koposov} S.~E.,   {Deason}
  A.~J.,  2018, \mn@doi [\mnras] {10.1093/mnras/sty982}, \href
  {https://ui.adsabs.harvard.edu/abs/2018MNRAS.478..611B} {478, 611}

\bibitem[\protect\citeauthoryear{{Belokurov}, {Sanders}, {Fattahi}, {Smith},
  {Deason}, {Evans}  \& {Grand}}{{Belokurov}
  et~al.}{2020}]{2020MNRAS.494.3880B}
{Belokurov} V.,  {Sanders} J.~L.,  {Fattahi} A.,  {Smith} M.~C.,  {Deason}
  A.~J.,  {Evans} N.~W.,   {Grand} R. J.~J.,  2020, \mn@doi [\mnras]
  {10.1093/mnras/staa876}, \href
  {https://ui.adsabs.harvard.edu/abs/2020MNRAS.494.3880B} {494, 3880}

\bibitem[\protect\citeauthoryear{{Bensby}, {Feltzing}  \&
  {Lundstr{\"o}m}}{{Bensby} et~al.}{2004}]{2004A&A...421..969B}
{Bensby} T.,  {Feltzing} S.,   {Lundstr{\"o}m} I.,  2004, \mn@doi [\aap]
  {10.1051/0004-6361:20035957}, \href
  {https://ui.adsabs.harvard.edu/abs/2004A&A...421..969B} {421, 969}

\bibitem[\protect\citeauthoryear{{Bergemann} et~al.,}{{Bergemann}
  et~al.}{2014}]{2014A&A...565A..89B}
{Bergemann} M.,  et~al., 2014, \mn@doi [\aap] {10.1051/0004-6361/201423456},
  \href {https://ui.adsabs.harvard.edu/abs/2014A&A...565A..89B} {565, A89}

\bibitem[\protect\citeauthoryear{{Bignone}, {Helmi}  \& {Tissera}}{{Bignone}
  et~al.}{2019}]{2019ApJ...883L...5B}
{Bignone} L.~A.,  {Helmi} A.,   {Tissera} P.~B.,  2019, \mn@doi [\apjl]
  {10.3847/2041-8213/ab3e0e}, \href
  {https://ui.adsabs.harvard.edu/abs/2019ApJ...883L...5B} {883, L5}

\bibitem[\protect\citeauthoryear{{Bland-Hawthorn} \&
  {Gerhard}}{{Bland-Hawthorn} \& {Gerhard}}{2016}]{2016ARA&A..54..529B}
{Bland-Hawthorn} J.,  {Gerhard} O.,  2016, \mn@doi [\araa]
  {10.1146/annurev-astro-081915-023441}, \href
  {https://ui.adsabs.harvard.edu/abs/2016ARA&A..54..529B} {54, 529}

\bibitem[\protect\citeauthoryear{{Bonaca} et~al.,}{{Bonaca}
  et~al.}{2020}]{2020ApJ...897L..18B}
{Bonaca} A.,  et~al., 2020, \mn@doi [\apjl] {10.3847/2041-8213/ab9caa}, \href
  {https://ui.adsabs.harvard.edu/abs/2020ApJ...897L..18B} {897, L18}

\bibitem[\protect\citeauthoryear{{Bovy}}{{Bovy}}{2015}]{2015ApJS..216...29B}
{Bovy} J.,  2015, \mn@doi [\apjs] {10.1088/0067-0049/216/2/29}, \href
  {https://ui.adsabs.harvard.edu/abs/2015ApJS..216...29B} {216, 29}

\bibitem[\protect\citeauthoryear{{Bressan}, {Marigo}, {Girardi}, {Salasnich},
  {Dal Cero}, {Rubele}  \& {Nanni}}{{Bressan}
  et~al.}{2012}]{2012MNRAS.427..127B}
{Bressan} A.,  {Marigo} P.,  {Girardi} L.,  {Salasnich} B.,  {Dal Cero} C.,
  {Rubele} S.,   {Nanni} A.,  2012, \mn@doi [\mnras]
  {10.1111/j.1365-2966.2012.21948.x}, \href
  {https://ui.adsabs.harvard.edu/abs/2012MNRAS.427..127B} {427, 127}

\bibitem[\protect\citeauthoryear{{Buck}}{{Buck}}{2020}]{2020MNRAS.491.5435B}
{Buck} T.,  2020, \mn@doi [\mnras] {10.1093/mnras/stz3289}, \href
  {https://ui.adsabs.harvard.edu/abs/2020MNRAS.491.5435B} {491, 5435}

\bibitem[\protect\citeauthoryear{{Buder} et~al.,}{{Buder}
  et~al.}{2021}]{2021MNRAS.506..150B}
{Buder} S.,  et~al., 2021, \mn@doi [\mnras] {10.1093/mnras/stab1242}, \href
  {https://ui.adsabs.harvard.edu/abs/2021MNRAS.506..150B} {506, 150}

\bibitem[\protect\citeauthoryear{{Casagrande}, {Sch{\"o}nrich}, {Asplund},
  {Cassisi}, {Ram{\'\i}rez}, {Mel{\'e}ndez}, {Bensby}  \&
  {Feltzing}}{{Casagrande} et~al.}{2011}]{2011A&A...530A.138C}
{Casagrande} L.,  {Sch{\"o}nrich} R.,  {Asplund} M.,  {Cassisi} S.,
  {Ram{\'\i}rez} I.,  {Mel{\'e}ndez} J.,  {Bensby} T.,   {Feltzing} S.,  2011,
  \mn@doi [\aap] {10.1051/0004-6361/201016276}, \href
  {https://ui.adsabs.harvard.edu/abs/2011A&A...530A.138C} {530, A138}

\bibitem[\protect\citeauthoryear{{Chiappini}, {Matteucci}  \&
  {Gratton}}{{Chiappini} et~al.}{1997}]{1997ApJ...477..765C}
{Chiappini} C.,  {Matteucci} F.,   {Gratton} R.,  1997, \mn@doi [\apj]
  {10.1086/303726}, \href
  {https://ui.adsabs.harvard.edu/abs/1997ApJ...477..765C} {477, 765}

\bibitem[\protect\citeauthoryear{{Ciuc{\u{a}}}, {Kawata}, {Miglio}, {Davies}
  \& {Grand}}{{Ciuc{\u{a}}} et~al.}{2021}]{2021MNRAS.503.2814C}
{Ciuc{\u{a}}} I.,  {Kawata} D.,  {Miglio} A.,  {Davies} G.~R.,   {Grand} R.
  J.~J.,  2021, \mn@doi [\mnras] {10.1093/mnras/stab639}, \href
  {https://ui.adsabs.harvard.edu/abs/2021MNRAS.503.2814C} {503, 2814}

\bibitem[\protect\citeauthoryear{{Clarke} et~al.,}{{Clarke}
  et~al.}{2019}]{2019MNRAS.484.3476C}
{Clarke} A.~J.,  et~al., 2019, \mn@doi [\mnras] {10.1093/mnras/stz104}, \href
  {https://ui.adsabs.harvard.edu/abs/2019MNRAS.484.3476C} {484, 3476}

\bibitem[\protect\citeauthoryear{{De Silva} et~al.,}{{De Silva}
  et~al.}{2015}]{2015MNRAS.449.2604D}
{De Silva} G.~M.,  et~al., 2015, \mn@doi [\mnras] {10.1093/mnras/stv327}, \href
  {https://ui.adsabs.harvard.edu/abs/2015MNRAS.449.2604D} {449, 2604}

\bibitem[\protect\citeauthoryear{{Donlon}, {Newberg}, {Sanderson}  \&
  {Widrow}}{{Donlon} et~al.}{2020}]{2020ApJ...902..119D}
{Donlon} Thomas I.,  {Newberg} H.~J.,  {Sanderson} R.,   {Widrow} L.~M.,  2020,
  \mn@doi [\apj] {10.3847/1538-4357/abb5f6}, \href
  {https://ui.adsabs.harvard.edu/abs/2020ApJ...902..119D} {902, 119}

\bibitem[\protect\citeauthoryear{{Dotter}, {Conroy}, {Cargile}  \&
  {Asplund}}{{Dotter} et~al.}{2017}]{2017ApJ...840...99D}
{Dotter} A.,  {Conroy} C.,  {Cargile} P.,   {Asplund} M.,  2017, \mn@doi [\apj]
  {10.3847/1538-4357/aa6d10}, \href
  {http://adsabs.harvard.edu/abs/2017ApJ...840...99D} {840, 99}

\bibitem[\protect\citeauthoryear{{Edvardsson}, {Andersen}, {Gustafsson},
  {Lambert}, {Nissen}  \& {Tomkin}}{{Edvardsson}
  et~al.}{1993}]{1993A&A...275..101E}
{Edvardsson} B.,  {Andersen} J.,  {Gustafsson} B.,  {Lambert} D.~L.,  {Nissen}
  P.~E.,   {Tomkin} J.,  1993, \aap, \href
  {https://ui.adsabs.harvard.edu/abs/1993A&A...275..101E} {500, 391}

\bibitem[\protect\citeauthoryear{{Feltzing}, {Holmberg}  \&
  {Hurley}}{{Feltzing} et~al.}{2001}]{2001A&A...377..911F}
{Feltzing} S.,  {Holmberg} J.,   {Hurley} J.~R.,  2001, \mn@doi [\aap]
  {10.1051/0004-6361:20011119}, \href
  {https://ui.adsabs.harvard.edu/abs/2001A&A...377..911F} {377, 911}

\bibitem[\protect\citeauthoryear{{Feuillet}, {Bovy}, {Holtzman}, {Girardi},
  {MacDonald}, {Majewski}  \& {Nidever}}{{Feuillet}
  et~al.}{2016}]{2016ApJ...817...40F}
{Feuillet} D.~K.,  {Bovy} J.,  {Holtzman} J.,  {Girardi} L.,  {MacDonald} N.,
  {Majewski} S.~R.,   {Nidever} D.~L.,  2016, \mn@doi [\apj]
  {10.3847/0004-637X/817/1/40}, \href
  {https://ui.adsabs.harvard.edu/abs/2016ApJ...817...40F} {817, 40}

\bibitem[\protect\citeauthoryear{{Feuillet} et~al.,}{{Feuillet}
  et~al.}{2018}]{2018MNRAS.477.2326F}
{Feuillet} D.~K.,  et~al., 2018, \mn@doi [\mnras] {10.1093/mnras/sty779}, \href
  {https://ui.adsabs.harvard.edu/abs/2018MNRAS.477.2326F} {477, 2326}

\bibitem[\protect\citeauthoryear{{Feuillet}, {Frankel}, {Lind}, {Frinchaboy},
  {Garc{\'\i}a-Hern{\'a}ndez}, {Lane}, {Nitschelm}  \&
  {Roman-Lopes}}{{Feuillet} et~al.}{2019}]{2019MNRAS.489.1742F}
{Feuillet} D.~K.,  {Frankel} N.,  {Lind} K.,  {Frinchaboy} P.~M.,
  {Garc{\'\i}a-Hern{\'a}ndez} D.~A.,  {Lane} R.~R.,  {Nitschelm} C.,
  {Roman-Lopes} A.,  2019, \mn@doi [\mnras] {10.1093/mnras/stz2221}, \href
  {https://ui.adsabs.harvard.edu/abs/2019MNRAS.489.1742F} {489, 1742}

\bibitem[\protect\citeauthoryear{{Gaia Collaboration} et~al.,}{{Gaia
  Collaboration} et~al.}{2016}]{2016A&A...595A...1G}
{Gaia Collaboration} et~al., 2016, \mn@doi [\aap]
  {10.1051/0004-6361/201629272}, \href
  {https://ui.adsabs.harvard.edu/abs/2016A&A...595A...1G} {595, A1}

\bibitem[\protect\citeauthoryear{{Gaia Collaboration} et~al.,}{{Gaia
  Collaboration} et~al.}{2021}]{2021A&A...649A...1G}
{Gaia Collaboration} et~al., 2021, \mn@doi [\aap]
  {10.1051/0004-6361/202039657}, \href
  {https://ui.adsabs.harvard.edu/abs/2021A&A...649A...1G} {649, A1}

\bibitem[\protect\citeauthoryear{{Gallart}, {Bernard}, {Brook}, {Ruiz-Lara},
  {Cassisi}, {Hill}  \& {Monelli}}{{Gallart}
  et~al.}{2019}]{2019NatAs...3..932G}
{Gallart} C.,  {Bernard} E.~J.,  {Brook} C.~B.,  {Ruiz-Lara} T.,  {Cassisi} S.,
   {Hill} V.,   {Monelli} M.,  2019, \mn@doi [Nature Astronomy]
  {10.1038/s41550-019-0829-5}, \href
  {https://ui.adsabs.harvard.edu/abs/2019NatAs...3..932G} {3, 932}

\bibitem[\protect\citeauthoryear{{Genovali} et~al.,}{{Genovali}
  et~al.}{2014}]{2014A&A...566A..37G}
{Genovali} K.,  et~al., 2014, \mn@doi [\aap] {10.1051/0004-6361/201323198},
  \href {https://ui.adsabs.harvard.edu/abs/2014A&A...566A..37G} {566, A37}

\bibitem[\protect\citeauthoryear{{Grand} et~al.,}{{Grand}
  et~al.}{2018}]{2018MNRAS.474.3629G}
{Grand} R. J.~J.,  et~al., 2018, \mn@doi [\mnras] {10.1093/mnras/stx3025},
  \href {https://ui.adsabs.harvard.edu/abs/2018MNRAS.474.3629G} {474, 3629}

\bibitem[\protect\citeauthoryear{{Hasselquist} et~al.,}{{Hasselquist}
  et~al.}{2019}]{2019ApJ...871..181H}
{Hasselquist} S.,  et~al., 2019, \mn@doi [\apj] {10.3847/1538-4357/aaf859},
  \href {https://ui.adsabs.harvard.edu/abs/2019ApJ...871..181H} {871, 181}

\bibitem[\protect\citeauthoryear{{Hayden}, {Recio-Blanco}, {de Laverny},
  {Mikolaitis}  \& {Worley}}{{Hayden} et~al.}{2017}]{2017A&A...608L...1H}
{Hayden} M.~R.,  {Recio-Blanco} A.,  {de Laverny} P.,  {Mikolaitis} S.,
  {Worley} C.~C.,  2017, \mn@doi [\aap] {10.1051/0004-6361/201731494}, \href
  {https://ui.adsabs.harvard.edu/abs/2017A&A...608L...1H} {608, L1}

\bibitem[\protect\citeauthoryear{{Haywood}, {Di Matteo}, {Lehnert}, {Katz}  \&
  {G{\'o}mez}}{{Haywood} et~al.}{2013}]{2013A&A...560A.109H}
{Haywood} M.,  {Di Matteo} P.,  {Lehnert} M.~D.,  {Katz} D.,   {G{\'o}mez} A.,
  2013, \mn@doi [\aap] {10.1051/0004-6361/201321397}, \href
  {https://ui.adsabs.harvard.edu/abs/2013A&A...560A.109H} {560, A109}

\bibitem[\protect\citeauthoryear{{Haywood}, {Lehnert}, {Di Matteo}, {Snaith},
  {Schultheis}, {Katz}  \& {G{\'o}mez}}{{Haywood}
  et~al.}{2016}]{2016A&A...589A..66H}
{Haywood} M.,  {Lehnert} M.~D.,  {Di Matteo} P.,  {Snaith} O.,  {Schultheis}
  M.,  {Katz} D.,   {G{\'o}mez} A.,  2016, \mn@doi [\aap]
  {10.1051/0004-6361/201527567}, \href
  {https://ui.adsabs.harvard.edu/abs/2016A&A...589A..66H} {589, A66}

\bibitem[\protect\citeauthoryear{{Haywood}, {Snaith}, {Lehnert}, {Di Matteo}
  \& {Khoperskov}}{{Haywood} et~al.}{2019}]{2019A&A...625A.105H}
{Haywood} M.,  {Snaith} O.,  {Lehnert} M.~D.,  {Di Matteo} P.,   {Khoperskov}
  S.,  2019, \mn@doi [\aap] {10.1051/0004-6361/201834155}, \href
  {https://ui.adsabs.harvard.edu/abs/2019A&A...625A.105H} {625, A105}

\bibitem[\protect\citeauthoryear{{Helmi}, {Babusiaux}, {Koppelman}, {Massari},
  {Veljanoski}  \& {Brown}}{{Helmi} et~al.}{2018}]{2018Natur.563...85H}
{Helmi} A.,  {Babusiaux} C.,  {Koppelman} H.~H.,  {Massari} D.,  {Veljanoski}
  J.,   {Brown} A. G.~A.,  2018, \mn@doi [\nat] {10.1038/s41586-018-0625-x},
  \href {https://ui.adsabs.harvard.edu/abs/2018Natur.563...85H} {563, 85}

\bibitem[\protect\citeauthoryear{{Hernandez}, {Valls-Gabaud}  \&
  {Gilmore}}{{Hernandez} et~al.}{1999}]{1999MNRAS.304..705H}
{Hernandez} X.,  {Valls-Gabaud} D.,   {Gilmore} G.,  1999, \mn@doi [\mnras]
  {10.1046/j.1365-8711.1999.02102.x}, \href
  {https://ui.adsabs.harvard.edu/abs/1999MNRAS.304..705H} {304, 705}

\bibitem[\protect\citeauthoryear{{Howes}, {Lindegren}, {Feltzing}, {Church}  \&
  {Bensby}}{{Howes} et~al.}{2019}]{2019A&A...622A..27H}
{Howes} L.~M.,  {Lindegren} L.,  {Feltzing} S.,  {Church} R.~P.,   {Bensby} T.,
   2019, \mn@doi [\aap] {10.1051/0004-6361/201833280}, \href
  {https://ui.adsabs.harvard.edu/abs/2019A&A...622A..27H} {622, A27}

\bibitem[\protect\citeauthoryear{{Isern}}{{Isern}}{2019}]{2019ApJ...878L..11I}
{Isern} J.,  2019, \mn@doi [\apjl] {10.3847/2041-8213/ab238e}, \href
  {https://ui.adsabs.harvard.edu/abs/2019ApJ...878L..11I} {878, L11}

\bibitem[\protect\citeauthoryear{{Jofr{\'e}}}{{Jofr{\'e}}}{2021}]{2021ApJ...920...23J}
{Jofr{\'e}} P.,  2021, \mn@doi [\apj] {10.3847/1538-4357/ac10c1}, \href
  {https://ui.adsabs.harvard.edu/abs/2021ApJ...920...23J} {920, 23}

\bibitem[\protect\citeauthoryear{{Johnson} et~al.,}{{Johnson}
  et~al.}{2021}]{2021arXiv210309838J}
{Johnson} J.~W.,  et~al., 2021, arXiv e-prints, \href
  {https://ui.adsabs.harvard.edu/abs/2021arXiv210309838J} {p. arXiv:2103.09838}

\bibitem[\protect\citeauthoryear{{J{\o}rgensen} \& {Lindegren}}{{J{\o}rgensen}
  \& {Lindegren}}{2005}]{2005ESASP.576..171J}
{J{\o}rgensen} B.~R.,  {Lindegren} L.,  2005, in {Turon} C.,  {O'Flaherty}
  K.~S.,   {Perryman} M.~A.~C.,  eds,  ESA Special Publication Vol. 576, The
  Three-Dimensional Universe with Gaia. p.~171

\bibitem[\protect\citeauthoryear{{Khoperskov}, {Haywood}, {Snaith}, {Di
  Matteo}, {Lehnert}, {Vasiliev}, {Naroenkov}  \& {Berczik}}{{Khoperskov}
  et~al.}{2021}]{2021MNRAS.501.5176K}
{Khoperskov} S.,  {Haywood} M.,  {Snaith} O.,  {Di Matteo} P.,  {Lehnert} M.,
  {Vasiliev} E.,  {Naroenkov} S.,   {Berczik} P.,  2021, \mn@doi [\mnras]
  {10.1093/mnras/staa3996}, \href
  {https://ui.adsabs.harvard.edu/abs/2021MNRAS.501.5176K} {501, 5176}

\bibitem[\protect\citeauthoryear{{Lebreton} \& {Goupil}}{{Lebreton} \&
  {Goupil}}{2014}]{2014A&A...569A..21L}
{Lebreton} Y.,  {Goupil} M.~J.,  2014, \mn@doi [\aap]
  {10.1051/0004-6361/201423797}, \href
  {https://ui.adsabs.harvard.edu/abs/2014A&A...569A..21L} {569, A21}

\bibitem[\protect\citeauthoryear{{Lindegren} et~al.,}{{Lindegren}
  et~al.}{2021}]{2021A&A...649A...2L}
{Lindegren} L.,  et~al., 2021, \mn@doi [\aap] {10.1051/0004-6361/202039709},
  \href {https://ui.adsabs.harvard.edu/abs/2021A&A...649A...2L} {649, A2}

\bibitem[\protect\citeauthoryear{{Martell} et~al.,}{{Martell}
  et~al.}{2017}]{2017MNRAS.465.3203M}
{Martell} S.~L.,  et~al., 2017, \mn@doi [\mnras] {10.1093/mnras/stw2835}, \href
  {https://ui.adsabs.harvard.edu/abs/2017MNRAS.465.3203M} {465, 3203}

\bibitem[\protect\citeauthoryear{{Martig} et~al.,}{{Martig}
  et~al.}{2016}]{2016MNRAS.456.3655M}
{Martig} M.,  et~al., 2016, \mn@doi [\mnras] {10.1093/mnras/stv2830}, \href
  {https://ui.adsabs.harvard.edu/abs/2016MNRAS.456.3655M} {456, 3655}

\bibitem[\protect\citeauthoryear{{Meibom} et~al.,}{{Meibom}
  et~al.}{2009}]{2009AJ....137.5086M}
{Meibom} S.,  et~al., 2009, \mn@doi [\aj] {10.1088/0004-6256/137/6/5086}, \href
  {https://ui.adsabs.harvard.edu/abs/2009AJ....137.5086M} {137, 5086}

\bibitem[\protect\citeauthoryear{{Michaud}, {Richard}, {Richer}  \&
  {VandenBerg}}{{Michaud} et~al.}{2004}]{2004ApJ...606..452M}
{Michaud} G.,  {Richard} O.,  {Richer} J.,   {VandenBerg} D.~A.,  2004, \mn@doi
  [\apj] {10.1086/383001}, \href
  {https://ui.adsabs.harvard.edu/abs/2004ApJ...606..452M} {606, 452}

\bibitem[\protect\citeauthoryear{{Mints}, {Hekker}  \& {Minchev}}{{Mints}
  et~al.}{2019}]{2019A&A...629A.127M}
{Mints} A.,  {Hekker} S.,   {Minchev} I.,  2019, \mn@doi [\aap]
  {10.1051/0004-6361/201935864}, \href
  {https://ui.adsabs.harvard.edu/abs/2019A&A...629A.127M} {629, A127}

\bibitem[\protect\citeauthoryear{{Mor}, {Robin}, {Figueras}, {Roca-F{\`a}brega}
   \& {Luri}}{{Mor} et~al.}{2019}]{2019A&A...624L...1M}
{Mor} R.,  {Robin} A.~C.,  {Figueras} F.,  {Roca-F{\`a}brega} S.,   {Luri} X.,
  2019, \mn@doi [\aap] {10.1051/0004-6361/201935105}, \href
  {https://ui.adsabs.harvard.edu/abs/2019A&A...624L...1M} {624, L1}

\bibitem[\protect\citeauthoryear{{Myeong}, {Vasiliev}, {Iorio}, {Evans}  \&
  {Belokurov}}{{Myeong} et~al.}{2019}]{2019MNRAS.488.1235M}
{Myeong} G.~C.,  {Vasiliev} E.,  {Iorio} G.,  {Evans} N.~W.,   {Belokurov} V.,
  2019, \mn@doi [\mnras] {10.1093/mnras/stz1770}, \href
  {https://ui.adsabs.harvard.edu/abs/2019MNRAS.488.1235M} {488, 1235}

\bibitem[\protect\citeauthoryear{{Naidu}, {Conroy}, {Bonaca}, {Johnson},
  {Ting}, {Caldwell}, {Zaritsky}  \& {Cargile}}{{Naidu}
  et~al.}{2020}]{2020ApJ...901...48N}
{Naidu} R.~P.,  {Conroy} C.,  {Bonaca} A.,  {Johnson} B.~D.,  {Ting} Y.-S.,
  {Caldwell} N.,  {Zaritsky} D.,   {Cargile} P.~A.,  2020, \mn@doi [\apj]
  {10.3847/1538-4357/abaef4}, \href
  {https://ui.adsabs.harvard.edu/abs/2020ApJ...901...48N} {901, 48}

\bibitem[\protect\citeauthoryear{{Nissen}, {Christensen-Dalsgaard},
  {Mosumgaard}, {Silva Aguirre}, {Spitoni}  \& {Verma}}{{Nissen}
  et~al.}{2020}]{2020A&A...640A..81N}
{Nissen} P.~E.,  {Christensen-Dalsgaard} J.,  {Mosumgaard} J.~R.,  {Silva
  Aguirre} V.,  {Spitoni} E.,   {Verma} K.,  2020, \mn@doi [\aap]
  {10.1051/0004-6361/202038300}, \href
  {https://ui.adsabs.harvard.edu/abs/2020A&A...640A..81N} {640, A81}

\bibitem[\protect\citeauthoryear{{Renaud}, {Agertz}, {Read}, {Ryde},
  {Andersson}, {Bensby}, {Rey}  \& {Feuillet}}{{Renaud}
  et~al.}{2021a}]{2021MNRAS.503.5846R}
{Renaud} F.,  {Agertz} O.,  {Read} J.~I.,  {Ryde} N.,  {Andersson} E.~P.,
  {Bensby} T.,  {Rey} M.~P.,   {Feuillet} D.~K.,  2021a, \mn@doi [\mnras]
  {10.1093/mnras/stab250}, \href
  {https://ui.adsabs.harvard.edu/abs/2021MNRAS.503.5846R} {503, 5846}

\bibitem[\protect\citeauthoryear{{Renaud}, {Agertz}, {Andersson}, {Read},
  {Ryde}, {Bensby}, {Rey}  \& {Feuillet}}{{Renaud}
  et~al.}{2021b}]{2021MNRAS.503.5868R}
{Renaud} F.,  {Agertz} O.,  {Andersson} E.~P.,  {Read} J.~I.,  {Ryde} N.,
  {Bensby} T.,  {Rey} M.~P.,   {Feuillet} D.~K.,  2021b, \mn@doi [\mnras]
  {10.1093/mnras/stab543}, \href
  {https://ui.adsabs.harvard.edu/abs/2021MNRAS.503.5868R} {503, 5868}

\bibitem[\protect\citeauthoryear{{Robin}, {Reyl{\'e}}, {Derri{\`e}re}  \&
  {Picaud}}{{Robin} et~al.}{2003}]{2003A&A...409..523R}
{Robin} A.~C.,  {Reyl{\'e}} C.,  {Derri{\`e}re} S.,   {Picaud} S.,  2003,
  \mn@doi [\aap] {10.1051/0004-6361:20031117}, \href
  {https://ui.adsabs.harvard.edu/abs/2003A&A...409..523R} {409, 523}

\bibitem[\protect\citeauthoryear{{Ruiz-Lara}, {Gallart}, {Bernard}  \&
  {Cassisi}}{{Ruiz-Lara} et~al.}{2020}]{2020NatAs...4..965R}
{Ruiz-Lara} T.,  {Gallart} C.,  {Bernard} E.~J.,   {Cassisi} S.,  2020, \mn@doi
  [Nature Astronomy] {10.1038/s41550-020-1097-0}, \href
  {https://ui.adsabs.harvard.edu/abs/2020NatAs...4..965R} {4, 965}

\bibitem[\protect\citeauthoryear{Sahlholdt}{Sahlholdt}{2020}]{samd_2020}
Sahlholdt C.~L.,  2020, csahlholdt/SAMD: First SAMD release,
  \mn@doi{10.5281/zenodo.3941733}

\bibitem[\protect\citeauthoryear{{Sahlholdt} \& {Lindegren}}{{Sahlholdt} \&
  {Lindegren}}{2021}]{2021MNRAS.502..845S}
{Sahlholdt} C.~L.,  {Lindegren} L.,  2021, \mn@doi [\mnras]
  {10.1093/mnras/stab034}, \href
  {https://ui.adsabs.harvard.edu/abs/2021MNRAS.502..845S} {502, 845}

\bibitem[\protect\citeauthoryear{{Sahlholdt}, {Casagrande}  \&
  {Feltzing}}{{Sahlholdt} et~al.}{2019}]{2019ApJ...881L..10S}
{Sahlholdt} C.~L.,  {Casagrande} L.,   {Feltzing} S.,  2019, \mn@doi [\apjl]
  {10.3847/2041-8213/ab321e}, \href
  {https://ui.adsabs.harvard.edu/abs/2019ApJ...881L..10S} {881, L10}

\bibitem[\protect\citeauthoryear{{Salaris}, {Chieffi}  \&
  {Straniero}}{{Salaris} et~al.}{1993}]{1993ApJ...414..580S}
{Salaris} M.,  {Chieffi} A.,   {Straniero} O.,  1993, \mn@doi [\apj]
  {10.1086/173105}, \href
  {https://ui.adsabs.harvard.edu/abs/1993ApJ...414..580S} {414, 580}

\bibitem[\protect\citeauthoryear{{Sellwood} \& {Binney}}{{Sellwood} \&
  {Binney}}{2002}]{2002MNRAS.336..785S}
{Sellwood} J.~A.,  {Binney} J.~J.,  2002, \mn@doi [\mnras]
  {10.1046/j.1365-8711.2002.05806.x}, \href
  {https://ui.adsabs.harvard.edu/abs/2002MNRAS.336..785S} {336, 785}

\bibitem[\protect\citeauthoryear{{Silva Aguirre} et~al.,}{{Silva Aguirre}
  et~al.}{2015}]{2015MNRAS.452.2127S}
{Silva Aguirre} V.,  et~al., 2015, \mn@doi [\mnras] {10.1093/mnras/stv1388},
  \href {https://ui.adsabs.harvard.edu/abs/2015MNRAS.452.2127S} {452, 2127}

\bibitem[\protect\citeauthoryear{{Silva Aguirre} et~al.,}{{Silva Aguirre}
  et~al.}{2018}]{2018MNRAS.475.5487S}
{Silva Aguirre} V.,  et~al., 2018, \mn@doi [\mnras] {10.1093/mnras/sty150},
  \href {https://ui.adsabs.harvard.edu/abs/2018MNRAS.475.5487S} {475, 5487}

\bibitem[\protect\citeauthoryear{{Skrutskie} et~al.,}{{Skrutskie}
  et~al.}{2006}]{2006AJ....131.1163S}
{Skrutskie} M.~F.,  et~al., 2006, \mn@doi [\aj] {10.1086/498708}, \href
  {https://ui.adsabs.harvard.edu/abs/2006AJ....131.1163S} {131, 1163}

\bibitem[\protect\citeauthoryear{{Snaith}, {Haywood}, {Di Matteo}, {Lehnert},
  {Combes}, {Katz}  \& {G{\'o}mez}}{{Snaith}
  et~al.}{2015}]{2015A&A...578A..87S}
{Snaith} O.,  {Haywood} M.,  {Di Matteo} P.,  {Lehnert} M.~D.,  {Combes} F.,
  {Katz} D.,   {G{\'o}mez} A.,  2015, \mn@doi [\aap]
  {10.1051/0004-6361/201424281}, \href
  {https://ui.adsabs.harvard.edu/abs/2015A&A...578A..87S} {578, A87}

\bibitem[\protect\citeauthoryear{{Soderblom}}{{Soderblom}}{2010}]{2010ARA&A..48..581S}
{Soderblom} D.~R.,  2010, \mn@doi [\araa]
  {10.1146/annurev-astro-081309-130806}, \href
  {https://ui.adsabs.harvard.edu/abs/2010ARA&A..48..581S} {48, 581}

\bibitem[\protect\citeauthoryear{{Spitoni}, {Silva Aguirre}, {Matteucci},
  {Calura}  \& {Grisoni}}{{Spitoni} et~al.}{2019}]{2019A&A...623A..60S}
{Spitoni} E.,  {Silva Aguirre} V.,  {Matteucci} F.,  {Calura} F.,   {Grisoni}
  V.,  2019, \mn@doi [\aap] {10.1051/0004-6361/201834188}, \href
  {https://ui.adsabs.harvard.edu/abs/2019A&A...623A..60S} {623, A60}

\bibitem[\protect\citeauthoryear{{Spitoni} et~al.,}{{Spitoni}
  et~al.}{2021}]{2021A&A...647A..73S}
{Spitoni} E.,  et~al., 2021, \mn@doi [\aap] {10.1051/0004-6361/202039864},
  \href {https://ui.adsabs.harvard.edu/abs/2021A&A...647A..73S} {647, A73}

\bibitem[\protect\citeauthoryear{{Valle}, {Dell'Omodarme}, {Prada Moroni}  \&
  {Degl'Innocenti}}{{Valle} et~al.}{2015}]{2015A&A...579A..59V}
{Valle} G.,  {Dell'Omodarme} M.,  {Prada Moroni} P.~G.,   {Degl'Innocenti} S.,
  2015, \mn@doi [\aap] {10.1051/0004-6361/201425568}, \href
  {https://ui.adsabs.harvard.edu/abs/2015A&A...579A..59V} {579, A59}

\bibitem[\protect\citeauthoryear{{Wu} et~al.,}{{Wu}
  et~al.}{2019}]{2019MNRAS.484.5315W}
{Wu} Y.,  et~al., 2019, \mn@doi [\mnras] {10.1093/mnras/stz256}, \href
  {https://ui.adsabs.harvard.edu/abs/2019MNRAS.484.5315W} {484, 5315}

\bibitem[\protect\citeauthoryear{{Zwitter} et~al.,}{{Zwitter}
  et~al.}{2020}]{2020arXiv201212201Z}
{Zwitter} T.,  et~al., 2020, arXiv e-prints, \href
  {https://ui.adsabs.harvard.edu/abs/2020arXiv201212201Z} {p. arXiv:2012.12201}

\makeatother
\end{thebibliography}




\appendix

\section{Choosing the regularization parameter} \label{app:choose_alpha}

\begin{figure*}
\centering
\includegraphics[width=\textwidth]{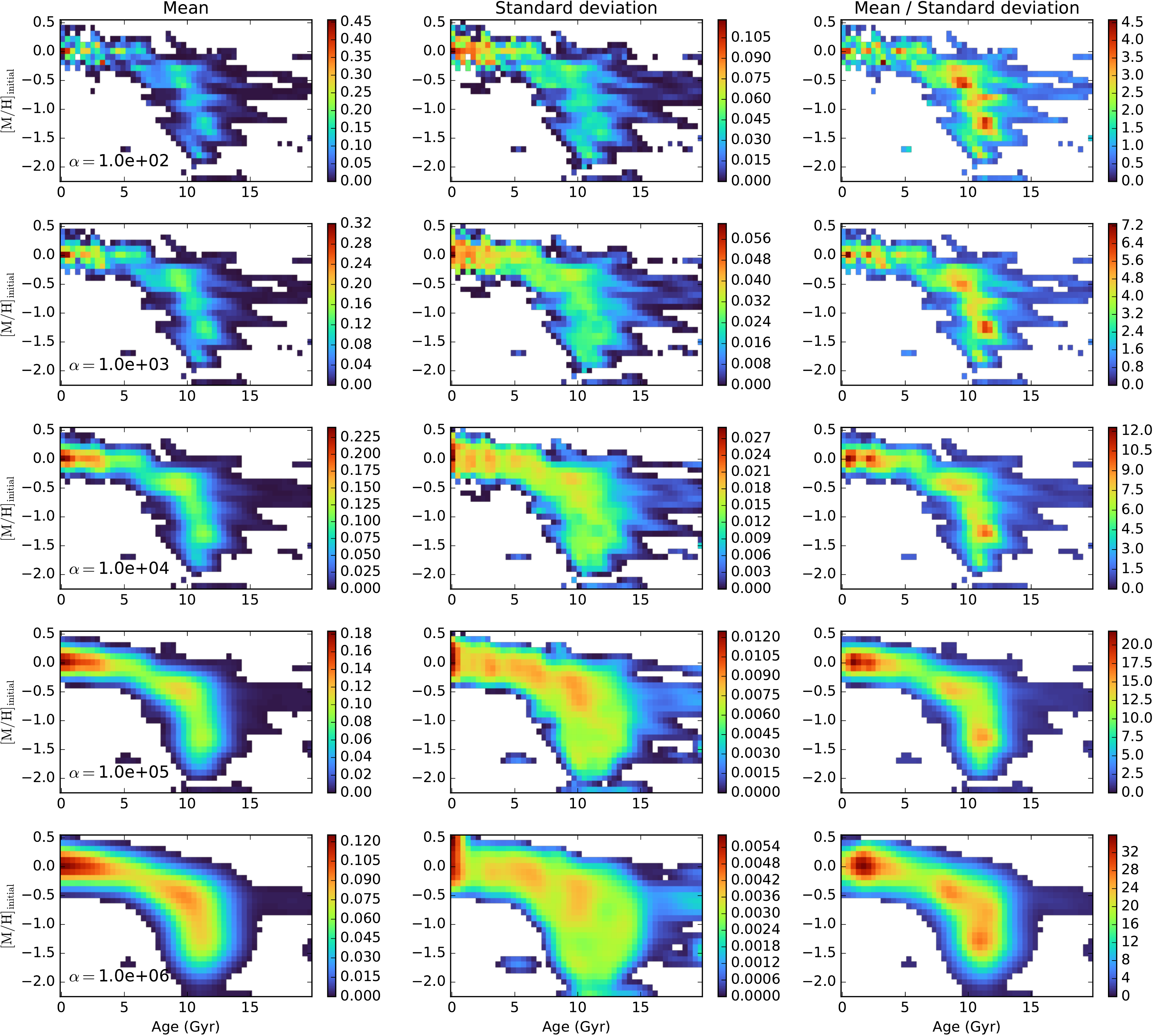}
\caption{SAMD calculated for a synthetic sample of 1600 dwarf and subgiants stars.
The first column shows the mean SAMD for five different values of $\alpha$ as indicated in the panels.
The mean is taken over 100 realizations of the SAMD calculated from random resamples of the data with replacement (i.e. each resample has the same size as the original sample, but duplicates are allowed).
The second and third columns show the standard deviation and mean divided by standard deviation (``S/N'') at each age-metallicity grid point for the same five $\alpha$-values.
As $\alpha$ increases, the mean becomes more smooth, the standard deviation decreases, and the S/N increases.
The true age-metallicity distribution of the sample is shown in \autoref{fig:SAMD_real_synth}.}
\label{fig:SAMD_mean_std_synth}
\end{figure*}

\begin{figure}
\centering
\includegraphics[width=\columnwidth]{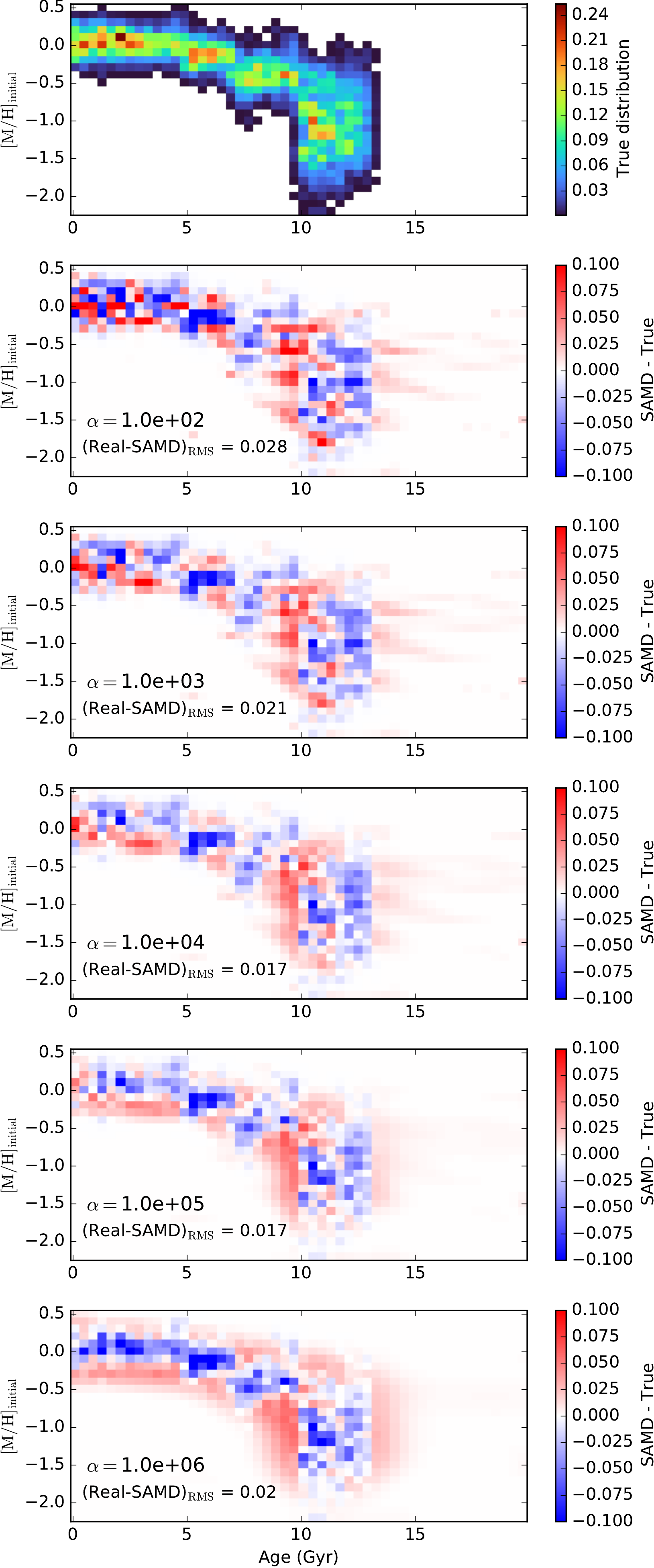}
\caption{Upper panel: the true distribution of ages and metallicities in the synthetic sample for which the SAMD is shown in \autoref{fig:SAMD_mean_std_synth}.
Remaining panels: the difference between the mean SAMD from \autoref{fig:SAMD_mean_std_synth} and the true age-metallicity distribution for five values of $\alpha$.
The lowest RMS difference is found for $\alpha = 10^{4}$ or $10^{5}$.}
\label{fig:SAMD_real_synth}
\end{figure}

\begin{figure}
\centering
\includegraphics[width=0.95\columnwidth]{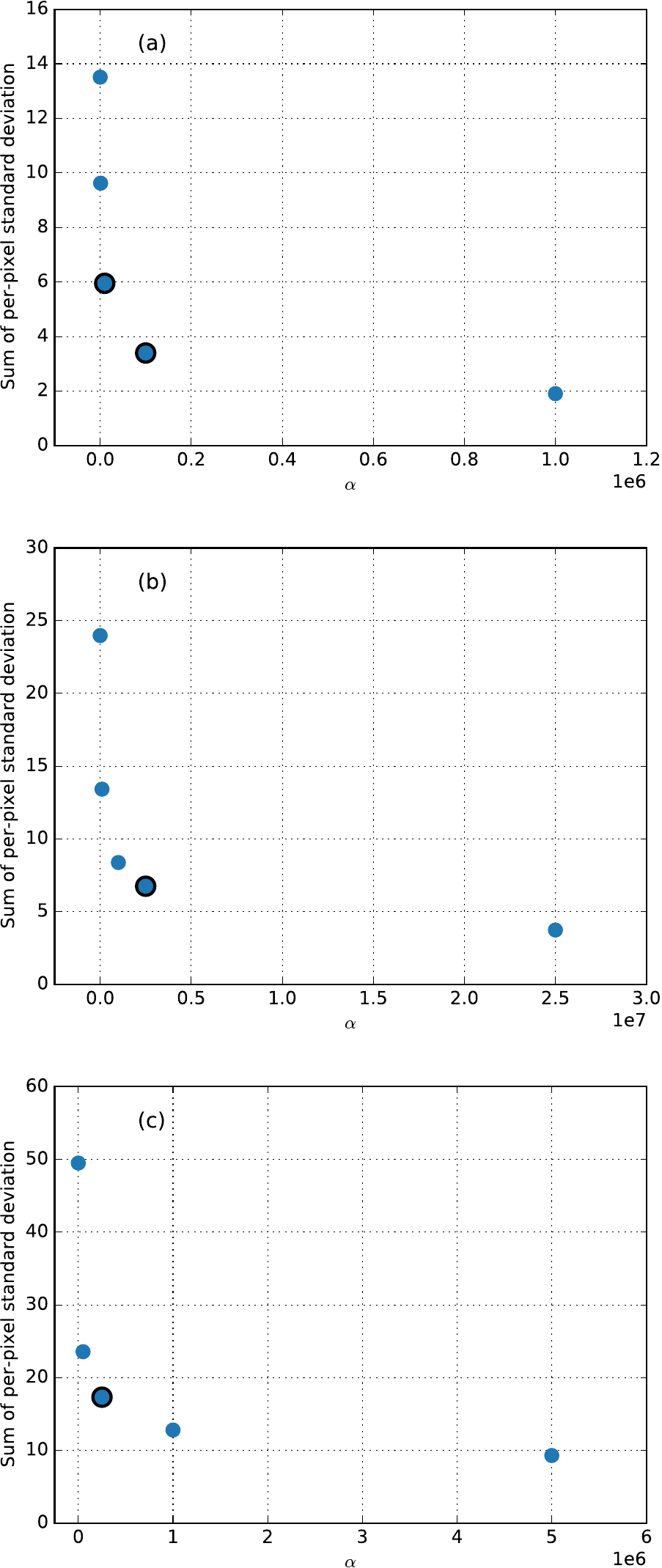}
\caption{(a) Sum of the standard deviation over all pixels of the SAMDs for the synthetic sample (shown in \autoref{fig:SAMD_mean_std_synth}) as a function of $\alpha$.
The standard deviation initially drops quickly when $\alpha$ is increased from $0$ but then levels off for high values of $\alpha$.
The optimal values of $\alpha$ are indicated by black rings around the markers.
These values are found close to the ``knee'' of the curve.
(b), (c) The same as panel a, but for the full GALAH sample (SAMDs shown in \autoref{fig:SAMD_mean_std}) and for the kinematically selected Population B (SAMDs shown in \autoref{fig:SAMD_mean_std_popb}).
The black ring marks the adopted value of $\alpha$.}
\label{fig:SAMD_std_alpha}
\end{figure}

\begin{figure*}
\centering
\includegraphics[width=\textwidth]{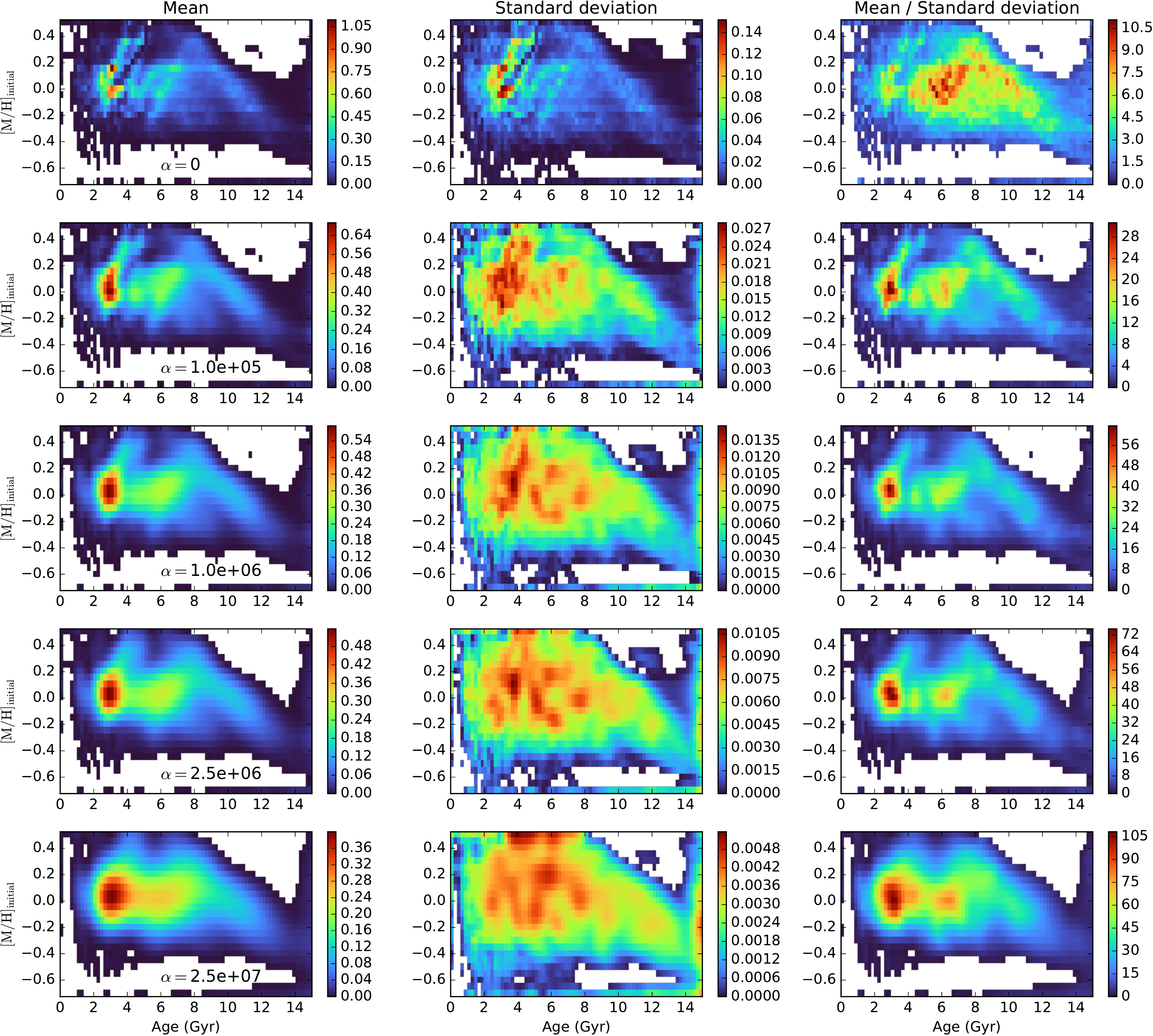}
\caption{The same as \autoref{fig:SAMD_mean_std_synth} but for the full GALAH sample.
Unlike the synthetic sample, the mean and standard deviation is taken over 40 realizations of the SAMD calculated from random resamples of 50\,000 stars \textit{without} replacement.}
\label{fig:SAMD_mean_std}
\end{figure*}


\begin{figure*}
\centering
\includegraphics[width=\textwidth]{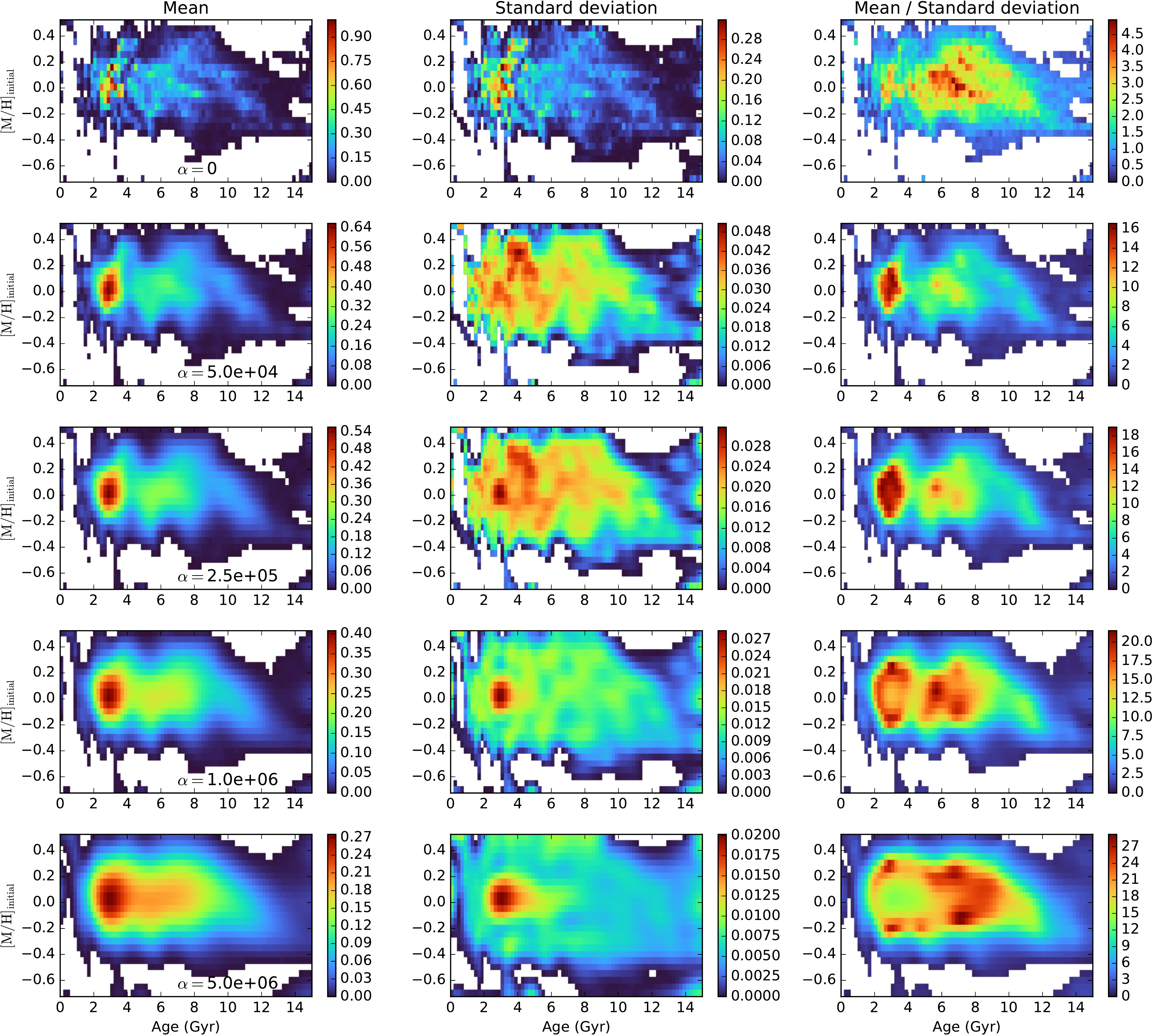}
\caption{The same as \autoref{fig:SAMD_mean_std_synth} but for Population B of the GALAH sample, i.e. the stars with $180$~km~s$^{-1} > V_{T} > 120$~km~s$^{-1}$.
The mean and standard deviation is taken over 40 realizations of the SAMD calculated from random resamples of 5\,000 stars \textit{without} replacement.}
\label{fig:SAMD_mean_std_popb}
\end{figure*}


As explained in Section~\ref{sec:method_SAMD}, we estimate the optimal value of the regularization parameter, $\alpha$ in Eq.~\eqref{eq:SAMD_eq}, using a resampling technique.
To justify and illustrate this method, we here present the SAMD for different values of $\alpha$ for samples of synthetic and real stars.

\citet{2021MNRAS.502..845S} tested the SAMD method on several synthetic samples of dwarfs and turnoff stars.
One of the samples was called the ``Milky Way-like'' sample because the age-metallicity distribution was defined to match the populations of the Besan\c{c}on Galaxy model \citep{2003A&A...409..523R}.
The MW-like sample contains 1600 stars.
\autoref{fig:SAMD_mean_std_synth} shows the mean and standard deviation of 100 realizations of the MW-like SAMD at five different values of $\alpha$.
The different realizations are calculated from random resamples of the data with the same sample size (1600 stars) allowing for duplicates.
\autoref{fig:SAMD_real_synth} shows the true age-metallicity distribution in the upper panel and the difference between the mean SAMD and the true distribution in the remaining panels.
The lowest RMS difference is found for $\alpha= 10^4$ and $10^5$.
For $\alpha$-values lower than this the difference looks more noisy (it shows large variations on short scales), and for the highest value of $\alpha = 10^6$, the SAMD underestimates the core of the distribution and overestimates the edges indicating that the solution has been smoothed too much.

Based on this test it seems that $\alpha$ in the range $10^4$ to $10^5$ is optimal for the MW-like sample.
However, for samples of real data we cannot make the comparison with the true distribution.
Therefore we consider how the sum over the standard deviation of the different SAMD realizations correlate with $\alpha$ as shown in \autoref{fig:SAMD_std_alpha}a.
We can think of the sum over the standard deviation as the noise in the solution.
As expected, there is a trade-off between noise and smoothness in the sense that the noise of the solution decreases as $\alpha$ and the smoothness increases.
In this case, the optimal value of $\alpha$ is around the ``knee'' of this trade-off curve.
The optimal solution also has a noise which is about 20 to 40 per cent of the low-$\alpha$ limit (i.e. no smoothing).

In addition to considering the noise alone, we can consider the mean divided by the standard deviation, or the signal-to-noise (S/N), which is shown in the third column of \autoref{fig:SAMD_mean_std_synth}.
For the MW-like sample, the S/N is fairly smooth and $\gtrsim 10$ over most of the distribution for $\alpha = 10^4$ to $10^5$.
This gives two indicators for whether the chosen value of $\alpha$ is reasonable:
Firstly, the noise of the solution, measured as the sum over the standard deviation of several random realizations of the SAMD, should be at the knee of the noise-smoothness tradeoff curve.
At that point the noise may be roughly 20 to 40 per cent of the low-$\alpha$ value.
Secondly, the S/N of the solution should be smooth and $\gtrsim 10$ over most of the distribution.
There is no guarantee that these criteria result in the optimal choice of $\alpha$ in every case, but they give a set of requirements which can be used to narrow down the range of $\alpha$.

In Figures~\ref{fig:SAMD_mean_std} and \ref{fig:SAMD_std_alpha}b the SAMD S/N and noise-smoothness tradeoff curve are shown for the full GALAH sample.
For this sample, the mean and standard deviation are calculated based on 40 samples of 50\,000 stars drawn randomly without duplicates.
Following the criteria given above, a value of $\alpha = 10^6$ to $2.5\times10^6$ seems reasonable.
For the final result presented in \autoref{fig:SAMD_all} we used a value of $\alpha=2.5\times10^6$.

For the smaller subsamples considered in sections~\ref{sec:vT_slice} and \ref{sec:Rz_slice} we select $\alpha$ based on a single subsample and apply the same value to every other subsample.
Figures~\ref{fig:SAMD_mean_std_popb} and \ref{fig:SAMD_std_alpha}c show the SAMD S/N and noise-smoothness tradeoff curve for Population B, i.e. the stars with $270$~km~s$^{-1} > V_{T} > 180$~km~s$^{-1}$.
For this sample, the mean and standard deviation are calculated based on 40 samples of 5\,000 stars drawn randomly without duplicates.
Based again on the criteria defined from the synthetic sample, we decide on a value of $\alpha=2.5\times10^5$.
We then calculate the SAMD for every subsample using this value and a maximum of 5\,000 stars.
This means that for subsamples with more than 5\,000 stars, we run 40 realizations of the SAMD with random draws of 5\,000 stars and take the mean.
For subsamples with fewer than 5\,000 stars we simply calculate the SAMD once for the entire sample.

\section{Low resolution SAMD} \label{app:lowres_SAMD}

\begin{figure*}
\centering
\includegraphics[width=\textwidth]{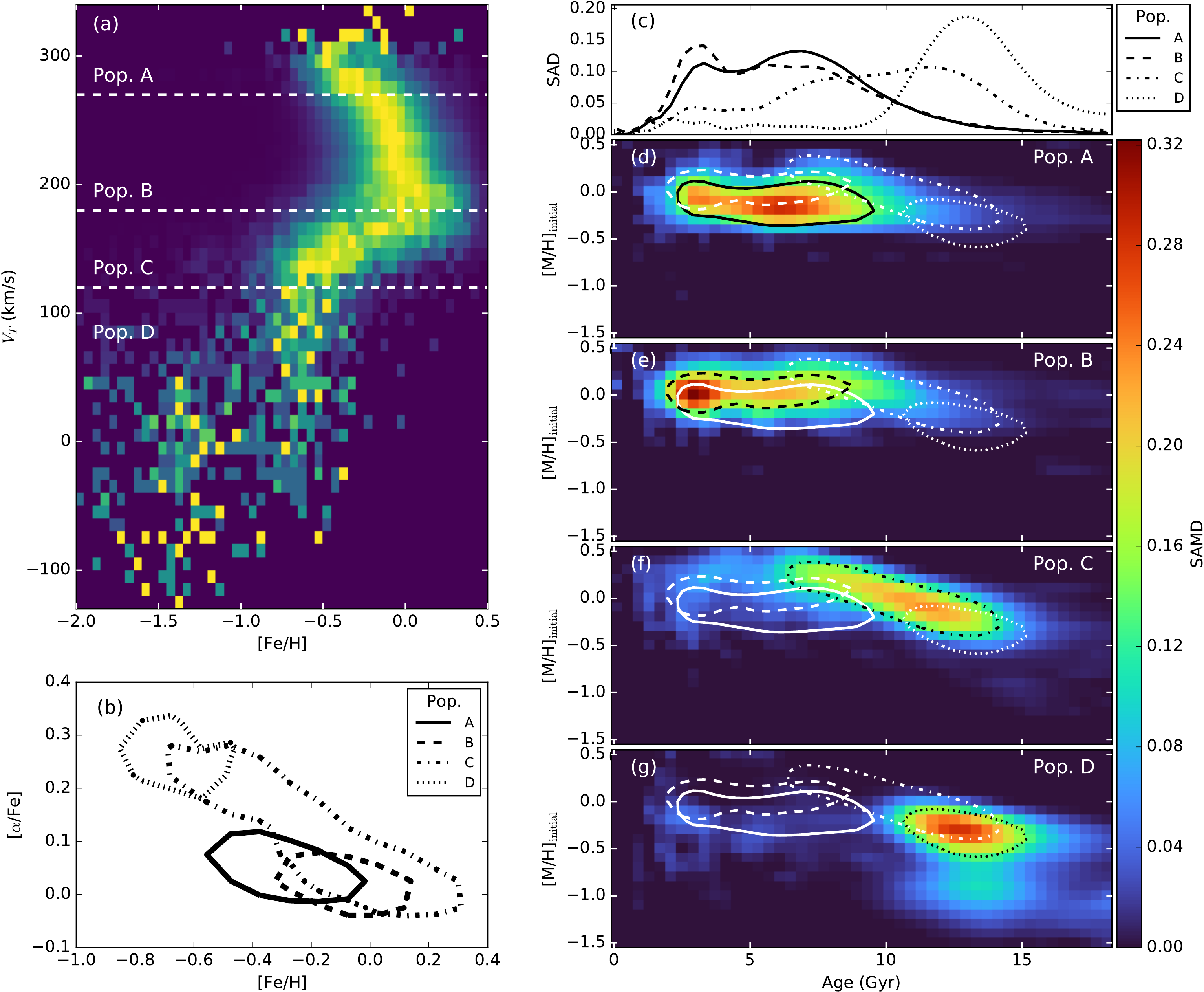}
\caption{Like \autoref{fig:SAMD_vT_bins} but with SAMDs calculated on a larger age-metallicity grid with half the resolution.}
\label{fig:SAMD_vT_bins_lowres}
\end{figure*}

\begin{figure*}
\centering
\includegraphics[width=\textwidth]{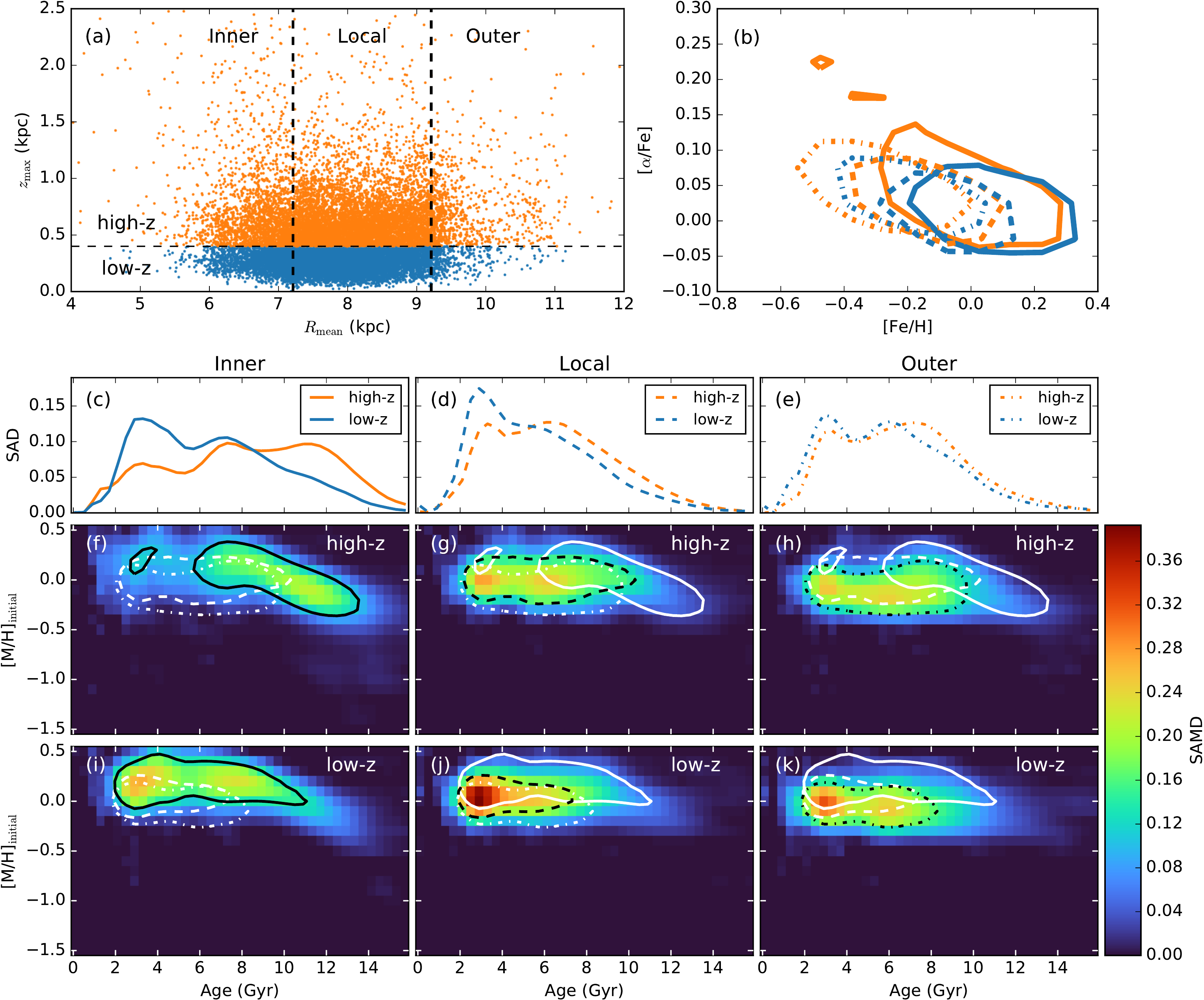}
\caption{Like \autoref{fig:SAMD_Rz_bins} but with SAMDs calculated on a larger age-metallicity grid with half the resolution.}
\label{fig:SAMD_Rz_bins_lowres}
\end{figure*}

Since some of the stars in the full sample fall outside of the high-resolution grid used to calculate the SAMD in Figures~\ref{fig:SAMD_vT_bins} and \ref{fig:SAMD_Rz_bins}, we show the results of calculating the SAMD on a larger grid with lower resolution in Figures~\ref{fig:SAMD_vT_bins_lowres} and \ref{fig:SAMD_Rz_bins_lowres}.
Although the impression of the distributions changes due to the resolution, only Population D in \autoref{fig:SAMD_vT_bins_lowres} shows features that fall outside the high-resolution grid.
This is because Population D is the only subsample with a significant fraction of stars at the oldest and most metal-poor edge of the sample.
We conclude that the use of the smaller high-resolution grid is justified for all populations except Population D.
Therefore, the combination of the high-resolution SAMDs for Populations A, B, and C in \autoref{fig:SAMD_vT_bins} and the low-resolution SAMD for Population D in \autoref{fig:SAMD_popD} is sufficient for a complete picture of the age-metallicity distributions of the kinematically selected subsamples.

\section{Correcting the SAMD for sample completeness} \label{app:SAMD_correction}

In section~\ref{sec:selection_function} we argued that the selection function of the sample, i.e. the magnitude and temperature limits, is unable to explain the local minima in the SAMD.
This is supported by the more or less monotonic behaviour of the age-metallicity completeness shown for different distances in \autoref{fig:completeness}.
Here we show what the SAMD of the full sample looks like after making a correction based on the sample completeness.

\begin{figure}
\centering
\includegraphics[width=\columnwidth]{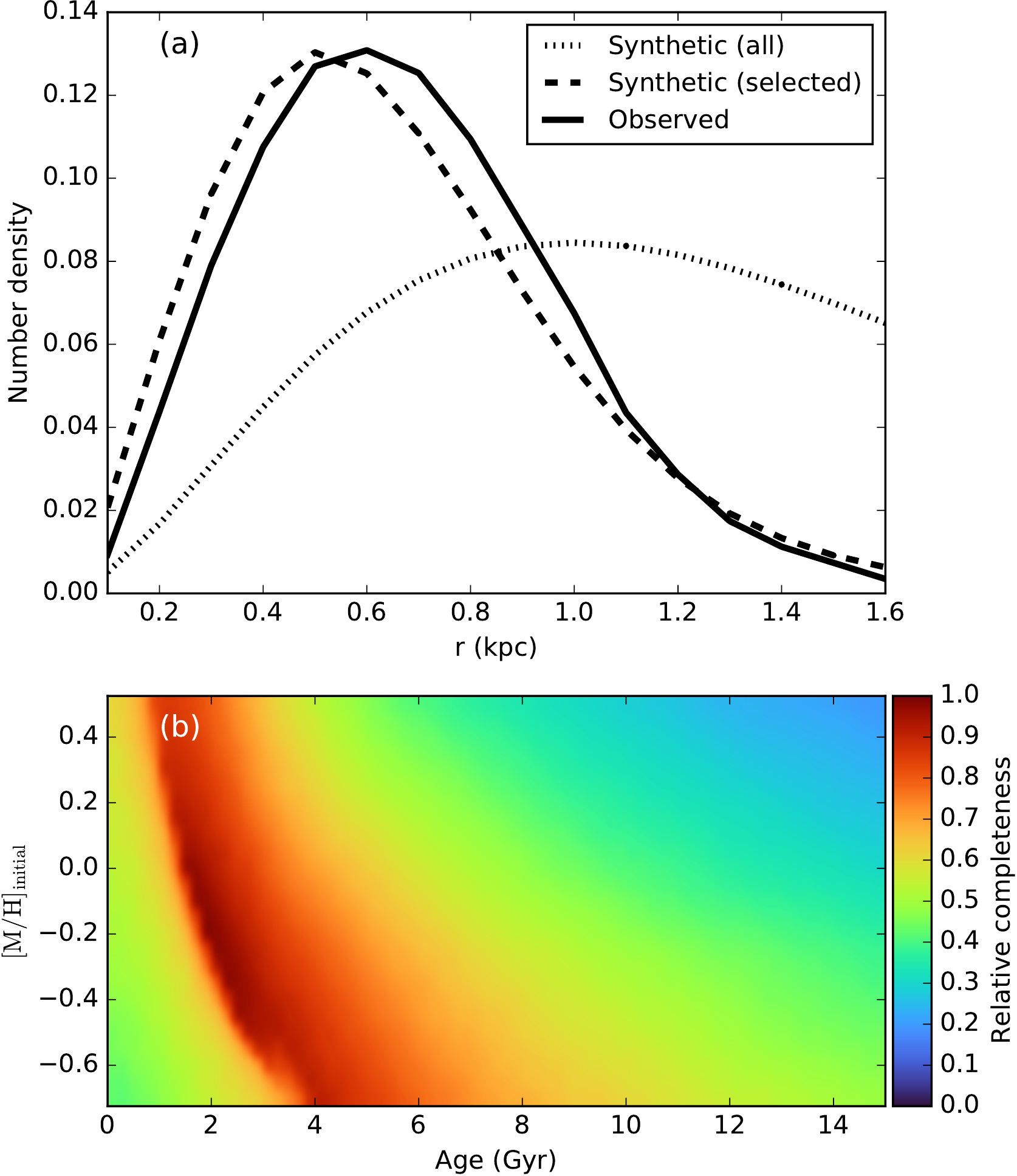}
\caption{(a) Distance distributions of the observed sample and the synthetic sample used to calculate the completeness function.
For the synthetic sample, the distribution is shown both before applying the selection function (given by Eq.~\eqref{eq:synth_dist}) and after.
(b). Age-metallicity completeness of the synthetic sample.}
\label{fig:SAMD_completeness}
\end{figure}

In order to define the completeness function of the sample as a whole, we must assume an underlying distance distribution.
Then the completeness function can be calculated as a linear combination of the completeness functions at different distances.
We make the assumption that the distance distribution is given by
\begin{equation} \label{eq:synth_dist}
    w(r) \propto r^2\exp(-2r) \; ,
\end{equation}
which is a constant density modulated by an exponential decrease.
With the distance in kpc, the exponential term has a scale length of $0.5$~kpc which is on the order of the scale height of the Galactic disc \citep{2016ARA&A..54..529B}.
For each distance in a range, $r = 0.1$ to $1.6$~kpc in steps of $0.1$~kpc, we define a synthetic sample and apply the GALAH selection function as described in section~\ref{sec:selection_function}.
At each distance, the size of the sample (before applying the selection), is proportional to $w(r)$.
This means that the complete synthetic sample is distributed in distance according to Eq.~\eqref{eq:synth_dist} as shown in \autoref{fig:SAMD_completeness}a.
\autoref{fig:SAMD_completeness}a also shows the distribution of distances in the synthetic sample after applying the selection function at each distance.
This distribution resembles the distance distribution of the observed sample fairly closely.
The completeness function calculated as the linear combination of the completeness functions at each distance is shown in \autoref{fig:SAMD_completeness}b.
Unsurprisingly, it looks like a mixture of the completeness functions at individual distances in \autoref{fig:completeness}.

\begin{figure}
\centering
\includegraphics[width=\columnwidth]{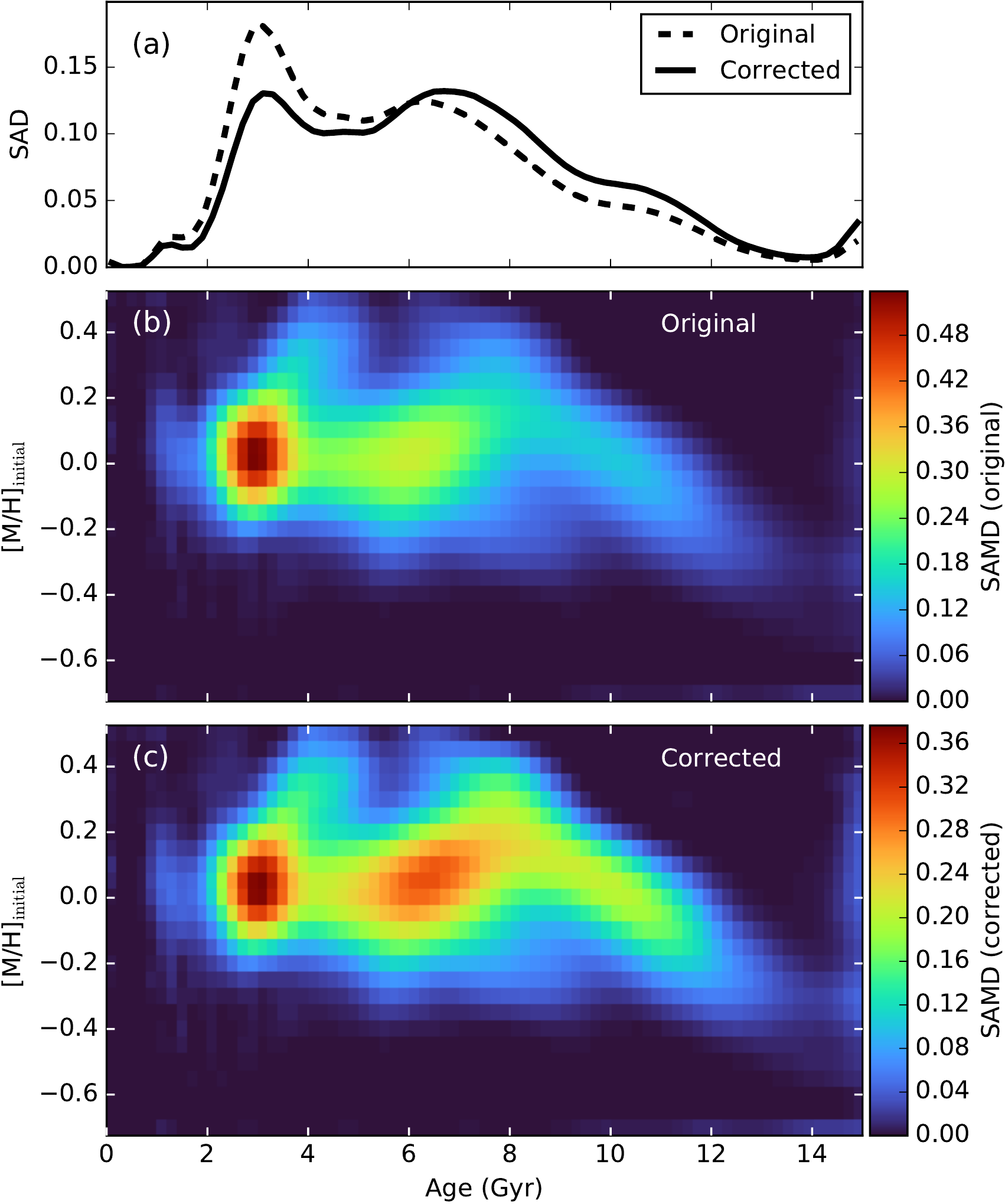}
\caption{(a) SAD for the full sample before and after correcting for sample completeness.
(b) SAMD for the full sample before correcting for sample completeness.
This is the same distribution as in \autoref{fig:SAMD_all}b.
(c) SAMD for the full sample after correcting for sample completeness.}
\label{fig:SAMD_corrected}
\end{figure}

Finally, by dividing the SAMD of the full sample with the sample completeness and renormalizing, we calculate the corrected SAMD shown in \autoref{fig:SAMD_corrected}.
As expected based on the discussion in section~\ref{sec:selection_function}, the main effect of this correction is to reduce the relative fraction of young stars.
At all ages below 6~Gyr, the corrected SAMD is lowered and vice versa for ages greater than 6~Gyr.
However, the correction does not significantly alter the shape of the SAMD, and it does not remove the local minima at 5~Gyr or the inflection at 10~Gyr.

\bsp	
\label{lastpage}
\end{document}